\documentclass[11pt]{article}
\usepackage{epsfig,sint,cite,amsfonts}
\usepackage{amssymb}
\usepackage{amsmath}
\usepackage[english]{babel}
\usepackage{simplewick}
\usepackage{bm}% bold math

\renewcommand{\vec}[1]{\boldsymbol{#1}}
\newcommand{\be}{\begin{equation}}
\newcommand{\ee}{\end{equation}}
\newcommand{\ba}{\begin{eqnarray}}
\newcommand{\ea}{\end{eqnarray}}
\newcommand{\la}{\label}

\newcommand{\<}{\langle}
\renewcommand{\>}{\rangle}

\newcommand{\txts}{\textstyle}

\begin{document}

$ $ \bigskip\bigskip\bigskip

\begin{center}
%{\LARGE\bf Thermal Current Correlators in $N_{\rm f}=2$ QCD\\[0.5ex]} 
% {\LARGE\bf Thermal Current Correlators, Electrical Conductivity and Dilepton Rates in $N_{\rm f}=2$ QCD\\[0.5ex]} 

% {\LARGE\bf Thermal Correlators in the $\rho$ channel \\[0.99ex]of two-flavor QCD} 
{\Large\bf Thermal Correlators in the $\rho$ channel of two-flavor QCD} 

\vskip 0.5cm

% {\large Erg\"anzung zur Vorlesung Theorie 5, Wintersemester 2011/12}
% \vskip 0.5cm

% {\large  Jun.-Prof.\ Harvey B.\ Meyer, JGU Mainz}
\end{center}
\vskip 1.5cm

%\begin{center}
% \noindent 
\centerline{\large Bastian B.\ Brandt$^{b,c}$, Anthony Francis$^{b,d}$, Harvey B.\ Meyer$^{a,b,d}$ % \\[0.5ex]
and Hartmut Wittig$^{a,b,d}$}
% \end{center}
\vskip 1.0cm
\noindent\hspace{-0.25cm} \emph{$^a$ PRISMA Cluster of Excellence, 
Johannes Gutenberg-Universit\"at Mainz, D-55099 Mainz}
\vskip 0.2cm
\noindent\hspace{-0.25cm} \emph{$^b$ Institut f\"ur Kernphysik, 
Johannes Gutenberg-Universit\"at Mainz, D-55099 Mainz}
\vskip 0.2cm
\noindent\hspace{-0.25cm} 
\emph{$^c$ Institut f\"ur theoretische Physik, Universit\"at Regensburg, D-93040 Regensburg}
\vskip 0.2cm
\noindent\hspace{-0.25cm} \emph{$^d$ Helmholtz Institut Mainz, Johannes Gutenberg-Universit\"at Mainz, D-55099 Mainz}

\vskip 2.5cm

\centerline{\bf Abstract}\medskip

\noindent We present a lattice QCD calculation with two dynamical flavors of the
isovector vector correlator in the high-temperature phase.  We analyze
the correlator in terms of the associated spectral function, for which
we review the theoretical expectations. In our main analysis, we
perform a fit for the difference of the thermal and vacuum spectral
functions, and we use an exact sum rule that constrains this
difference. We also perform a direct fit for the thermal spectral
function, and obtain good agreement between the two analyses for
frequencies below the two-pion threshold. Under the assumption that
the spectral function is smooth in that region, we give an estimate of
the electrical conductivity.

\vspace{1.5cm}


\thispagestyle{empty}

\newpage

\clearpage
\pagenumbering{arabic} 
%%%%%%%%%%%%%%%%%%%%%%%%%%%%%%%%%%%%%%%%%%%%%%%%%%%%%%%%%%%%%%%%%%%%%%%%
%%%%%%%%%%%%%%%%%%%%%%%%%%%%%%%%%%%%%%%%%%%%%%%%%%%%%%%%%%%%%%%%%%%%%%%%
%%%%%%%%%%%%%%%%%%%%%%%%%%%%%%%%%%%%%%%%%%%%%%%%%%%%%%%%%%%%%%%%%%%%%%%%
\section{Introduction}

The properties of strongly interacting matter under extreme conditions
are the subject of intensive experimental and theoretical
investigation. A comprehensive picture of a state of matter requires
not only the knowledge of equilibrium properties such as the equation
of state and static susceptibilities, but also an understanding of its
transport properties.

In the high-temperature phase of QCD, the transport coefficients (the
shear and bulk viscosities as well as the electrical conductivity) have been
calculated perturbatively to full leading order in the strong coupling
$\alpha_s$~\cite{Arnold:2001ms,Arnold:2003zc,Arnold:2006fz}. However,
in the range of temperatures that can be reached in heavy ion
collisions, the perturbative uncertainty remains large.
Phenomenologically, the observation of large elliptic flow in heavy
ion collisions at RHIC and at the LHC hints at a small shear viscosity
($\!\!$\cite{Teaney:2009qa,Song:2012tv} and references
therein). Furthermore, the measured spectrum of dileptons is related to
the spectral functions of the electromagnetic current, integrated over
the history of the expanding system~\cite{Rapp:1999ej}.

Any reliable calculation of a transport coefficient of QCD at a
temperature of a few hundred MeV would be extremely valuable. It
therefore makes sense to undertake a calculation in the comparatively
easiest possible channel. For a lattice QCD approach, the spectral
function of the isovector vector current is perhaps the most
accessible channel. First, the correlation function is evaluated via a
single, connected Wick contraction, allowing for a good signal-to-noise 
ratio in the Monte-Carlo simulation. Second, the light quarks do
not introduce a new dynamical scale into the problem in the way that
heavy quarks do, which typically requires the use of an effective
field theory. Third, the corresponding vacuum spectral function, being
extremely well known experimentally due to decades of measurements of
the $R$ ratio and of $\tau$ decays (see for
instance~\cite{Hagiwara:2011af}), provides a useful reference.

Here we present the first lattice calculation of the Euclidean
isovector vector correlator in the high-temperature phase of QCD with
dynamical quark flavors and analyze it in terms of the spectral
function.  We adopt an approach used previously in the bulk
channel~\cite{Meyer:2010ii}, which consists in analyzing directly the
difference of the thermal and vacuum correlators. Moreover, we exploit
a recently derived sum rule which constrains the integral over the
difference of spectral functions (divided by frequency) to vanish. We
compare our results at finite lattice spacing with a recent analysis
performed in the continuum limit of quenched QCD~\cite{Ding:2010ga}.

In spite of the technically favorable properties of the channel,
determining the vector spectral function with frequency resolution
$\Delta\omega\ll T$ remains a numerically ill-posed problem (see for
instance the discussion in~\cite{Meyer:2011gj}). Our main goal in this
paper is therefore of qualitative nature and consists in determining
the gross features of the thermal spectral function, in particular in
which frequency bins (of width $\Delta\omega\gtrsim 2T$) it under- or
overshoots the vacuum spectral function.

%%%%%%%%%%%%%%%%%%%%%%%%%%%%%%%%%%%%%%%%%%%%%%%%%%%%%%%%%%%%%%%%%%%%%%%%
%%%%%%%%%%%%%%%%%%%%%%%%%%%%%%%%%%%%%%%%%%%%%%%%%%%%%%%%%%%%%%%%%%%%%%%%
%%%%%%%%%%%%%%%%%%%%%%%%%%%%%%%%%%%%%%%%%%%%%%%%%%%%%%%%%%%%%%%%%%%%%%%%
\section{Theoretical expectations for the spectral function}

% Transport review: \cite{Meyer:2011gj}
% Quenched vector correlators \cite{Ding:2010ga}
% CLS scale setting: \cite{Fritzsch:2012wq}
% MPHMC code \cite{Marinkovic:2010eg}
% DDHMC code \cite{CLScode}
% Basti's runs \cite{Brandt:2010bn}, \cite{Brandt:2010uw}

In this section we set up our notation and define the relevant correlation functions.
We summarize the theoretical expectations for these correlators and the associated 
transport properties.

\subsection{Definitions}

Our primary observables are the Euclidean vector current correlators,
\ba
G_{\mu\nu}(\tau,T) &=&   \int d^3x \; \< J_\mu(\tau,\vec x)\;J_\nu(0)^\dagger\>\,,
\ea
with $J_\mu(x)\equiv \frac{1}{\sqrt{2}}\left(\bar u(x)\gamma_\mu u(x) - \bar
d(x)\gamma_\mu d(x)\right)$ the isospin current.  The expectation values are taken
with respect to the equilibrium density matrix $e^{-\beta H}/Z(\beta)$, where
$\beta\equiv 1/T$ is the inverse temperature.
The quark number susceptibility is defined as
\be
\chi_s\equiv -\int d^4x\; \< J_0(x) J_0(0)\> = -\beta \int d^3x \;\< J_0(\tau,\vec x) J_0(0)\>.
\label{eq:chis}
\ee
Due to charge conservation, the two correlators of interest are exactly related via
\be
G_{ii}(\tau,T)=\chi_s T + G_{\mu\mu}(\tau,T)~~.
\label{eq:GiGv}
\ee
The Euclidean correlators have the spectral representation
\be
G_{\mu\nu}(\tau,T)  = \int_0^\infty \frac{d\omega}{2\pi} \; \rho_{\mu\nu}(\omega,T) \;
\frac{\cosh[\omega(\beta/2-\tau)]}{\sinh(\omega\beta/2)}\;.
\ee
For a given function $\rho(\omega,T)$, the reconstructed correlator is defined as
\be 
G^{\rm rec}(\tau,T;T') {\equiv} \int_0^\infty \frac{d\omega}{2\pi}\; \rho(\omega,T')\;
\frac{\cosh[\omega(\frac{\beta}{2}-\tau)]}{\sinh( \omega \beta/2)} \,.
\la{eq:Grec1-main}
\ee
It can be interpreted as the Euclidean correlator that would be
realized at temperature $T$ if the spectral function was unchanged between
temperature $T$ and $T'$.  For $T'=0$ it can be directly obtained from the
zero-temperature Euclidean correlator via~\cite{Meyer:2010ii}
\be
G^{\rm rec}(\tau,T) \equiv G^{\rm rec}(\tau,T;0) = 
\sum_{m\in\mathbb{ Z}} G(|\tau+m\beta|,T=0).
\la{eq:Grec2-main}
\ee

\subsection{Theoretical predictions}

The isospin diffusion constant $D$ is given by a Kubo formula in terms of the
low-frequency behavior of the spectral function,
\be
D\chi_s = \frac{1}{6}\lim_{\omega\to0} \frac{\rho_{ii}(\omega,T)}{\omega}.
\la{eq:kubo}
\ee

In the thermodynamic limit, the subtracted vector spectral function obeys a sum rule
(see~\cite{Bernecker:2011gh} sec.\ 3.2),
\be\la{eq:sr}
\int_{-\infty}^\infty \frac{d\omega}{\omega} \; \Delta\rho(\omega,T) = 0,
\qquad \qquad \Delta\rho(\omega,T) \equiv  \rho_{ii}(\omega,T)-\rho_{ii}(\omega,0).
\ee
This sum rule is based on two ingredients. Firstly, the two-point function of a
spatial component of the vector current at vanishing four-momentum can
be interpreted as the susceptibility of the isospin charge at zero
temperature in a system with one short spatial periodic dimension of
length $\beta$.  As long as the correlation lengths are finite, this
susceptibility vanishes. Secondly, subtracting the same quantity in the 
infinite-volume vacuum enables one to write a convergent sum rule in $\omega$
for the susceptibility. It is well known from the operator-product
expansion that the difference of spectral functions in
Eq.\ (\ref{eq:sr}) falls off as $1/\omega^2$ at large frequencies, see
\cite{Chetyrkin:1985kn,Mallik:1997pq} for explicit calculations.

For non-interacting massive quarks in the fundamental representation of the SU($N_c$) color group, 
the vector spectral function is diagonal in flavor space and takes the form
\ba\la{eq:SFfree}
\rho_{ii}(\omega,T) &=&  2\pi \chi_s \<v^2\>\omega \delta(\omega) 
 \\ && 
+ \frac{N_c}{2\pi} \theta(\omega-2m) \left[1-\Big(\frac{2m}{\omega}\Big)^2\right]
\left[1+ {\txts\frac{1}{2}} \Big(\frac{2m}{\omega}\Big)^2 \right] \omega^2 \tanh\frac{\omega}{4T}.
\nonumber
\ea
The next-to-leading order has been computed very recently~\cite{Burnier:2012ze}.
At large frequencies the radiative corrections
$(1+\alpha_s/\pi+\dots)$ to the coefficient of the $\omega^2$ term are
temperature independent and known to 
order $\alpha_s^4$~\cite{Baikov:2012zm} (for quark mass effects in the vacuum,
see~\cite{Chetyrkin:1997qi}). The susceptibility and the mean squared transport
velocity $\<v^2\>$ have the following expressions,
\ba
\chi_s &=& {4N_c\beta}\int \frac{d^3\vec p}{(2\pi)^3} \; f_{\vec p} (1-f_{\vec p})
\\
\chi_s \<v^2\> &=& {4N_c\beta}\int \frac{d^3\vec p}{(2\pi)^3}\; f_{\vec p} (1-f_{\vec p})\;
               \frac{\vec p^2}{E_{\vec p}^2}
\ea
with $f_{\vec p}= 1/[e^{\beta E_{\vec p}}+1]$ the Fermi-Dirac distribution and $E_{\vec p} =
\sqrt{\vec p^2+m^2}$.  It is now straightforward to check that the sum
rule (\ref{eq:sr}) is verified in the free theory.  The positive
contribution of the transport peak (the $\omega\delta(\omega)$ contribution 
in the free theory) is compensated by a deficit of the thermal spectral function 
at intermediate frequencies, $\omega=\mathcal{O}(T)$ in the massless case.
The susceptibility and mean square velocity have
simple expressions in the massless and in the heavy-quark limits,

\be\begin{tabular}{l|l|l}
   &   $m=0$  &  $m \gg T$ \\
\hline
$ \chi_s$   & $\frac{N_c}{3}T^2$  &  $\frac{4N_c}{T}\Big[\frac{mT}{2\pi}\Big]^{{3}/{2}}\,e^{-m/T}  \phantom{\Bigg(\Bigg)}$ \\
$ \<v^2\>$  &  1                  & ${3T}/{m} \;.$ 
\end{tabular}
\ee
% \ba
% \chi_s,~~~\<v^2\> &=& \left\{ \begin{array}{l@{\qquad}l@{\qquad}l} 
% \frac{N_c}{3}T^2, & 1  & m=0  \\
% \frac{4N_c}{T}\Big[\frac{mT}{2\pi}\Big]^{\frac{3}{2}}\,e^{-\beta m} ,
% & \frac{3T}{m}  & m \gg T.
% \end{array}\right.~~~~
% \ea

Beyond the non-interacting theory, at weak coupling kinetic theory
predicts the presence of a narrow transport peak in the spectral
function at $\omega=0$, whose width and height are related to the
properties of the quasi-particles. Introducing a separation scale
$\Lambda$ between the transport time scale and the thermal time-scale,
the area under the transport peak is, to leading order, preserved by
the interactions~\cite{Petreczky:2005nh},
\be\la{eq:IntLbda}
\mathcal{A}(\Lambda)=\int_{-\Lambda}^\Lambda \frac{d\omega}{2\pi} \frac{\rho_{ii}(\omega,T)}{\omega} = 
\chi_s \<v^2\>.
\ee
The width of the transport peak however becomes finite. In the
heavy-quark limit for instance, the Langevin effective theory predicts
a Lorentzian form~\cite{Petreczky:2005nh}
\be\la{eq:rhoHQ}
\rho_{ii}(\omega,T) = \chi_s \<v^2\> \frac{2\eta\,\omega}{\omega^2+\eta^2},
\qquad \qquad m\gg T,
\ee
where $\eta$ is the `drag coefficient', $1/\eta= D\frac{m}{T}$.  The
case of massless quarks can be treated with the Boltzmann
equation~\cite{Arnold:2001ms,Moore:2006qn,Hong:2010at}, with a form of
the spectral function qualitatively similar to Eq.~(\ref{eq:rhoHQ})
and a drag coefficient given by
\be
\eta =\frac{g^2}{8\pi}C_Fm_D^2\log(T/m_D),
\ee
with $m_D$ the Debye mass and $C_F=\frac{N_c^2-1}{2N_c}$.

Finally, in contrast with the weak-coupling analysis outlined above,
it is worth mentioning that at least one theory is known where the
vector spectral function does not exhibit a transport peak. In the
strongly coupled ${\cal N}=4$ super-Yang-Mills theory, the spectral
function of the R-charge correlator reads~\cite{Myers:2007we},
\be
\la{eq:N4vc}
\rho^R_{ii}(\omega,T) = \frac{3\chi_s}{2\pi} \,\left(\frac{\omega}{ T}\right)^2
\frac{\sinh\frac{\omega}{2T}}{\cosh\frac{\omega}{2T}-\cos\frac{\omega}{2T}}\,.
\ee
The static susceptibility is given by $\chi_s= \frac{N_c^2T^2}{8}$, and the 
diffusion constant by $D=\frac{1}{2\pi T}$.

\subsection{Connection with electromagnetic observables}
\label{sec:pheno}

The electrical conductivity $\sigma$ is extracted from the correlator
of the current $J_\mu^{\rm em} = \sum_f Q_f \bar q_f \gamma_\mu
q_f$. If the quark species are degenerate, this correlator is given by
the sum of a contribution proportional to $\sum_f Q_f^2$ and a
contribution proportional to $(\sum_f Q_f)^2$.  At high frequencies,
the latter contribution to the spectral function is predicted to be
small in perturbative QCD. Assuming that this contribution is also
small at low frequencies, we have for the electrical conductivity in
$N_{\rm f}=2$ QCD
\be
\la{eq:sigD}
\sigma = C_{em} D\chi_s.
\ee
with $C_{em} = \sum_{f=u,d} Q_f^2=5/9$.  If one further assumes that
$\sigma$ is fairly insensitive to the quark mass, and that the
transport properties are not significantly affected by the presence of
virtual strange quark pairs, the electrical conductivity in $N_f=2+1$ QCD 
is given by Eq.~(\ref{eq:sigD}) with $C_{em} =\sum_{f=u,d,s} Q_f^2=2/3$.

The phenomenological significance of the electromagnetic spectral
function is that the dilepton rate and the real-photon production rate
are given by~\cite{McLerran:1984ay,Moore:2006qn} 
\ba
{{\rm d} N_{l^+l^-} \over {\rm d}\omega {\rm d}^3p} &=&
C_{em}{\alpha^2_{em}  \over 6 \pi^3} {\rho_{\mu\mu}(\omega,\vec{p},T) 
\over (\omega^2-\vec{p}^2) ({\rm e}^{\omega/T} - 1)}
\label{eq:dilrate}
\quad ,\\
\lim_{\omega \rightarrow 0} \omega \frac{{\rm d} R_\gamma}{{\rm d}^3p} &=&
\frac{3}{2\pi^2} \sigma(T) T \alpha_{em} \,
\ea
where $\alpha_{em}$ is the electromagnetic fine structure constant.

\subsection{Phenomenology of the $\rho$ resonance\la{sec:phenorho}}

In the vacuum, the QCD spectral function of the electromagnetic
current is well measured via the $R(s=\omega^2)$ ratio. The $\rho$
meson completely dominates the spectral function up to about
$\omega=1{\rm GeV}$. We work in the exact isospin symmetric theory and
therefore ignore issues related to isospin breaking and $\rho-\omega$
mixing, see~\cite{Jegerlehner:2011ti} for a recent reference.

Since we are working with the isospin current, we should restrict
ourselves to final hadronic states with $I=1$. The $R_1(s)$ ratio is
defined analogously to $R(s)$ with this restriction. A rough
parametrization of the experimentally measured $R_1(s)$ ratio was
given in Ref.\ \cite{Bernecker:2011gh}, Eq.\ (93). The spectral function in
our normalization is related to the $R_1$ ratio via
\be
\rho_{ii}(\omega,0) = \frac{1}{\pi}R_1(\omega^2)\,\omega^2.
\ee
We can now make a simple argument about the thermal spectral function 
based on the exact sum rule (\ref{eq:sr}),
the kinetic theory sum rule (\ref{eq:IntLbda}) and 
the experimentally known vacuum spectral function.
Using the parametrization of~\cite{Bernecker:2011gh},
the area under the vacuum spectral function up to $\omega=1{\rm GeV}$ is about
\be\la{eq:AreaRho}
{\cal A}_\rho \equiv \int_{0}^{1.0{\rm GeV}}
\frac{d\omega}{\pi}\frac{\rho_{ii}(\omega,0)}{\omega} \approx 0.114{\rm GeV}^2.
\ee
By contrast, the corresponding area for free massless quarks at zero temperature is
\be
{\cal A}_{\rm free} = 0.076{\rm GeV}^2.
\ee
Taking into account the physical light quark masses changes this value 
by a negligeable amount. 

\begin{figure}[t]
\centerline{
\includegraphics[width=.55\textwidth]{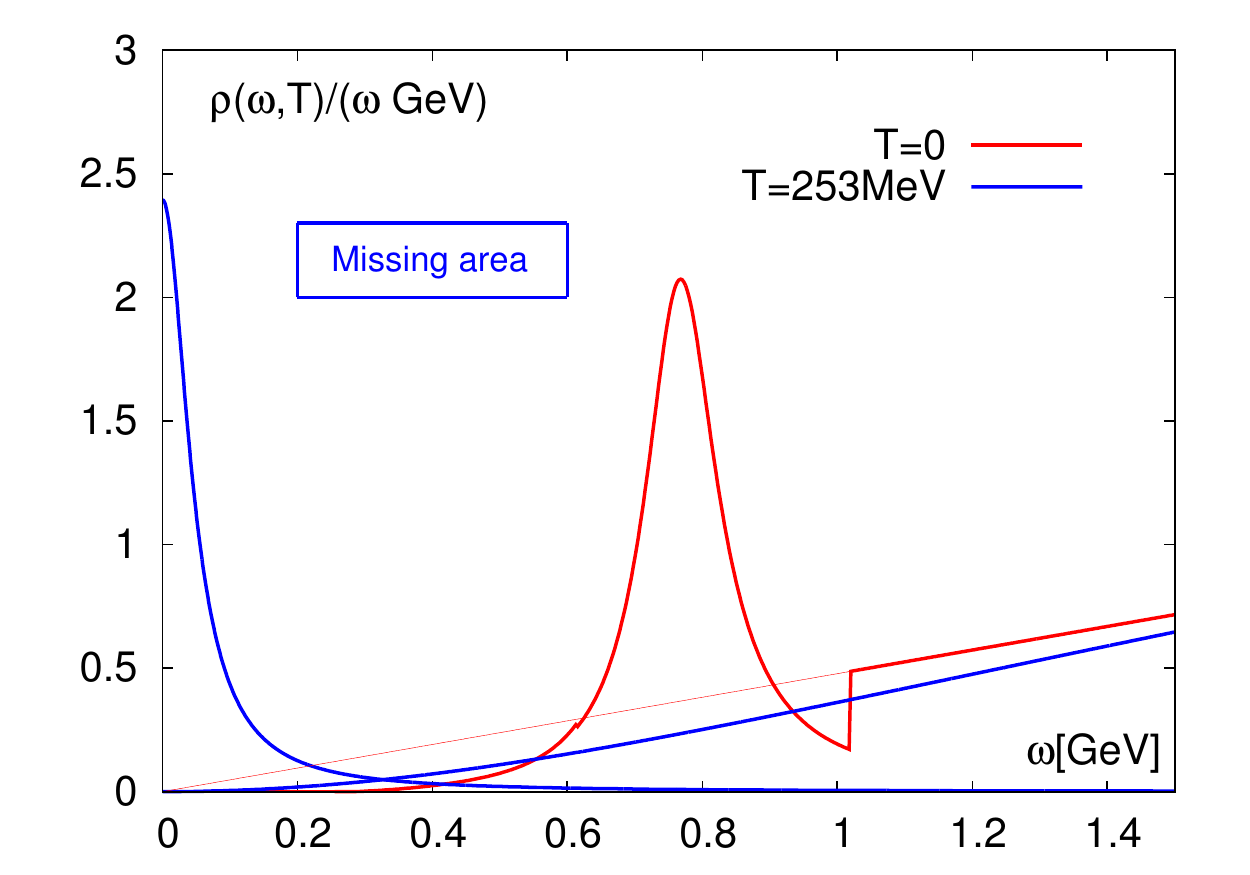}\hspace{-0.2cm}
\includegraphics[width=.55\textwidth]{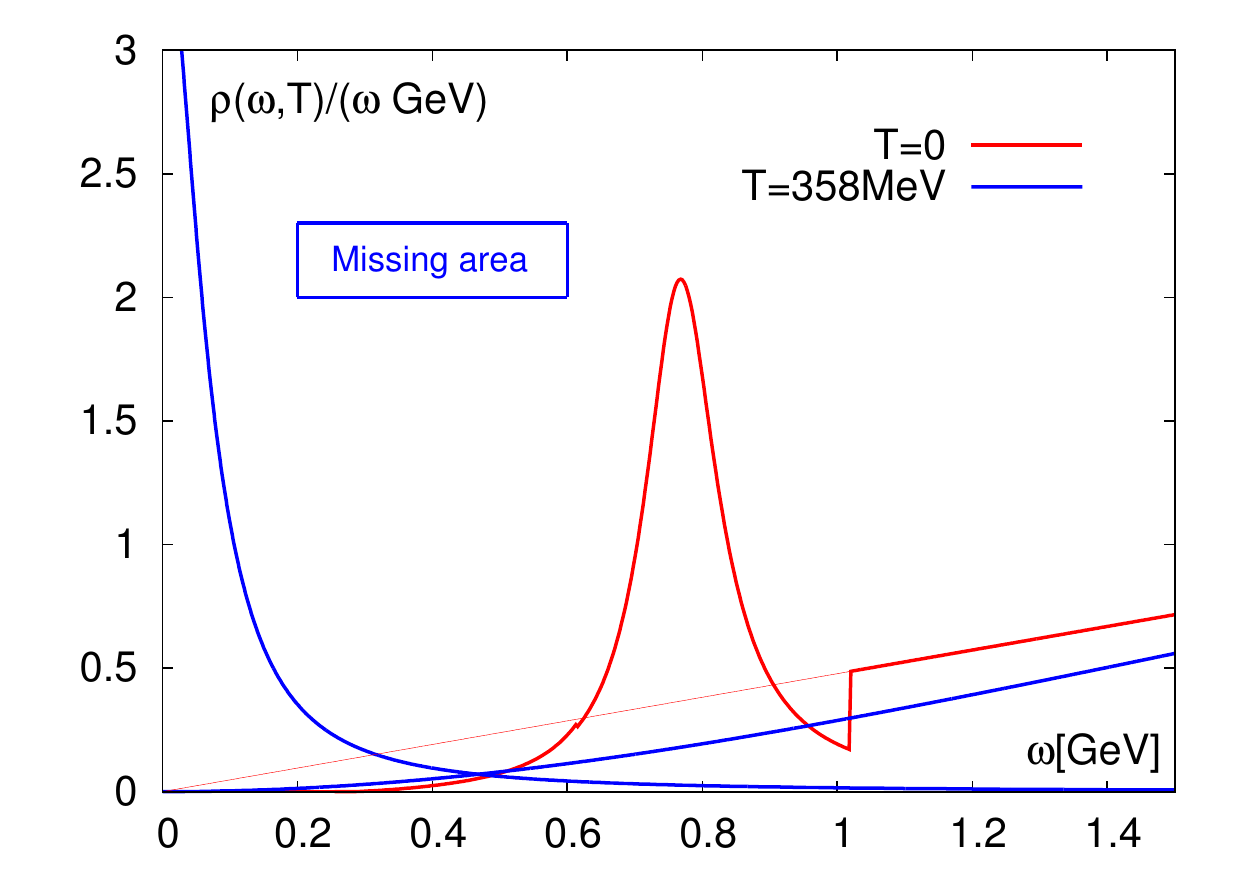}}
\caption{\it\small The phenomenological isovector vector spectral function $\rho_{ii}$ in the vacuum
compared to the leading-order weak-coupling prediction for the spectral function
in the high-temperature phase. The Born term is as given in Eq.\ (\ref{eq:SFfree}) for $m=0$ and 
we have represented the transport peak around $\omega=0$
by a Lorentzian with a width matched to \cite{Moore:2006qn} for 
$\alpha_s=0.3$ and an area determined by $\<v^2\>=1$ together with $\chi_s/\chi_s^{_{\rm SB}}=0.88$ and $0.9$
respectively for the left and right panel (table \ref{tab:pars} and \cite{Borsanyi:2011sw}). 
The thin red line represents the vacuum spectral function
for non-interacting massless quarks. 
The weak-coupling thermal spectral functions must receive additional contributions of the size 
given by the `missing area' rectangle in order to satisfy the sum rule (\ref{eq:sr}).
}
\label{fig:rho0}
\end{figure}

In the free theory, the sum rule (\ref{eq:sr}) is satisfied. Now
switching on interactions between quarks, let us assume for the sake
of the argument that in the high temperature phase they can be
described perturbatively. In the small frequency region, interactions
turn the delta function in Eq.\ (\ref{eq:SFfree}) into an approximate
Lorentzian curve~\cite{Moore:2006qn}, \emph{preserving its area to
  leading order}. This is the content of a kinetic theory sum
rule~\cite{Petreczky:2005nh}. In the vacuum, interactions have a
dramatic effect on the spectral function, due to chiral symmetry
breaking and confinement, and convert its area up to 1GeV from ${\cal
  A}_{\rm free}$ to ${\cal A}_{\rm \rho}$. Since the sum rule
(\ref{eq:sr}) must still be satisfied, the weak-coupling spectral
function in the high-temperature phase must acquire an additional area
of
\be\la{eq:Amiss}
{\cal A}_{\rm missing} = {\cal A}_\rho - {\cal A}_{\rm free} \approx 0.038{\rm GeV}^2.
\ee
Note that at very high temperatures, the area (\ref{eq:Amiss}) is
negligible compared to the area $\mathcal{A}(\Lambda)=\chi_s
\<v^2\>$ under the transport peak, which grows as $T^2$. However,
consider the situation at $T=253{\rm MeV }$ (the temperature at which
 we have computed the correlator
on the lattice, see the next section).  Assuming the existence
of a transport peak, its area $\mathcal{A}(\Lambda)$ is about
$0.056{\rm GeV}^2$, if we correct for the fact that the static isospin
susceptibility is about $12\%$ below its Stefan-Boltzmann limit $\chi_s^{_{\rm SB}}$ (see
table \ref{tab:pars} and \cite{Borsanyi:2011sw}).  The area missing
from the weak-coupling spectral function is thus comparable in size to
the transport peak area at this temperature.  The argument is
illustrated in Fig.\ (\ref{fig:rho0}).

% \footnote{In the figure we also corrected for
%   the fact that the static isospin susceptibility is about $12\%$
%   below its Stefan-Boltzmann limit for the temperatures considered
%   (see table \ref{tab:pars} and \cite{Borsanyi:2011sw}).}.

At $\omega\gg T$, the difference of spectral functions
$\Delta\rho(\omega,T)$ can be analyzed with the operator product
expansion (see~\cite{Mallik:1997pq,CaronHuot:2009ns} and References
therein). The lowest-dimensional gauge-invariant operators are of
dimension four, and therefore the asymptotic behavior is
$\Delta\rho(\omega,T)\propto 1/\omega^2$ (possibly up to logarithms).
According to Eq.\ (4.1) of Ref.\ \cite{CaronHuot:2009ns}, the
leading term of order $1/\omega^2$ is positive. However, at $T=253{\rm
  MeV}$ its contribution to the area under
$\Delta\rho(\omega,T)/\omega$ is too small to explain the missing area
(\ref{eq:Amiss}).

In conclusion, at temperatures that are accessible in heavy-ion
collisions, the sum rule (\ref{eq:sr}) and the $R(s)$ ratio
measurements place an important constraint on the thermal spectral
function.

%%%%%%%%%%%%%%%%%%%%%%%%%%%%%%%%%%%%%%%%%%%%%%%%%%%%%%%%%%%%%%%%%%%%%%%%
%%%%%%%%%%%%%%%%%%%%%%%%%%%%%%%%%%%%%%%%%%%%%%%%%%%%%%%%%%%%%%%%%%%%%%%%
%%%%%%%%%%%%%%%%%%%%%%%%%%%%%%%%%%%%%%%%%%%%%%%%%%%%%%%%%%%%%%%%%%%%%%%%
\section{Lattice QCD data}

All our numerical results were computed on dynamical gauge
configurations with two light, mass-degenerate
$\mathcal{O}(a)$-improved Wilson quark flavors. The configurations
were generated using the MP-HMC
algorithm~\cite{Hasenbusch:2001ne,Hasenbusch:2002ai} in the
implementation of Marinkovic and Schaefer~\cite{Marinkovic:2010eg}
based on L\"uscher's DD-HMC package~\cite{CLScode}.  The
zero-temperature ensemble was made available to us through the CLS
effort~\cite{CLS}. The gauge action is the standard Wilson plaquette
action \cite{Wilson:1974sk}, while the fermions were implemented via
the Wilson-Clover discretization with non-perturbatively determined
clover coefficient $c_{\rm sw}$ \cite{Jansen:1998mx}.

We calculated correlation functions using the same discretization and
masses as in the sea sector in two different ensembles. The first
corresponds to virtually zero-temperature on a $64^3\times 128$
lattice (labeled O7 in~\cite{Fritzsch:2012wq}) with a lattice spacing
of $a=0.0486(4)(5)$fm~\cite{Fritzsch:2012wq} and a pion mass of
$m_\pi=270$MeV, so that $m_\pi L = 4.2$.  Secondly we generated an
ensemble on a lattice of size $64^3\times 16$ with all bare parameters
identical to the zero-temperature ensemble.  In this way it is
straightforward to compare the correlation functions respectively in
the confined and deconfined phases of QCD.  Choosing $N_\tau=16$
yields a temperature of $T\simeq250$MeV. Based on preliminary results
on the pseudo-critical temperature $T_c$ of the crossover from the
hadronic to the high-temperature phase~\cite{Brandt:2012sk}, the
temperature can also be expressed as $T/T_c\approx 1.2$.

The renormalization of the vector correlator assumes the form 
\be
G_{\mu\nu}(\tau,\beta,g_0)=Z_V^2(g_0)[1+b_V(g_0)am_q]^2
\left( G_{\mu\nu}^{lat}(\tau,\beta,g_0) + \dots\right).
\ee
The dots refer to O($a$) contributions from the improvement term
proportional to the derivative of the antisymmetric tensor
operator~\cite{Luscher:1996sc,Sint:1997jx} that we did not compute. We
note however that perturbatively, the O($a$) contribution is
suppressed by two powers of~$\alpha_s$. We use the non-perturbative
value of $Z_V$ provided by~\cite{DellaMorte:2005rd}. From the
perturbative results of \cite{Sint:1997jx} we estimated the magnitude
of the term $\sim b_V(g_0)am_q$ to be of the order of
$0.3\%$. Considering that the error of $Z_V^2$ is at the $1\%$ level,
we drop this contribution altogether in the following. Note that since
we use the same bare parameters at zero and finite temperature, the
multiplicative renormalization of the vector current only affects the
corresponding correlation functions -- and their difference -- by an
overall factor.

\begin{figure}[t]
\centering
\includegraphics[width=.63\textwidth]{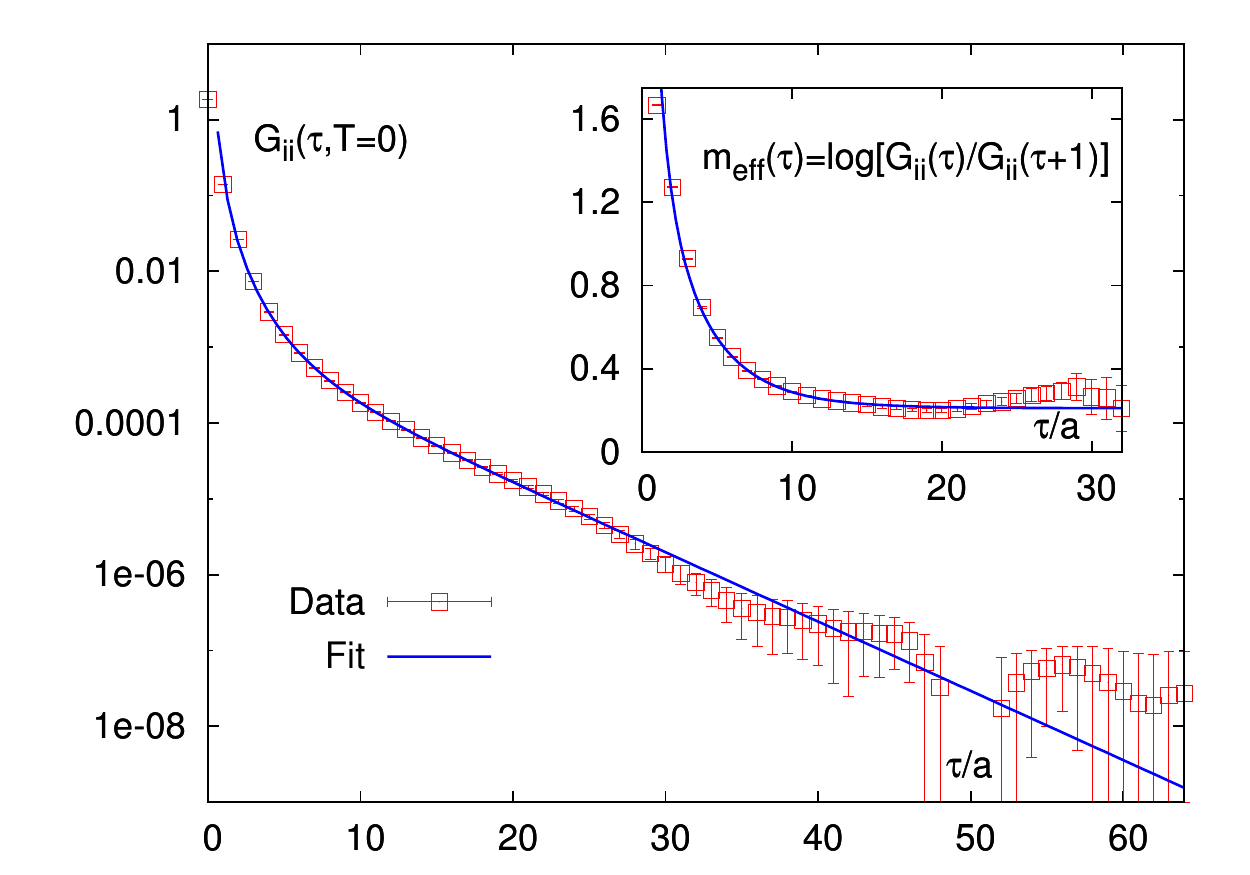}
\caption{\it{The vacuum vector correlator $G_{ii}(\tau,T=0)$ computed on a
  lattice sized $64^3\times 128$ and $m_\pi=270$MeV, labeled O7 in
  \cite{Fritzsch:2012wq}. The blue line denotes the fit result using
  Eq.~\ref{eqn:efmfit}. The insertion shows the corresponding
  effective mass for $0\leq\tau/a\leq32$.}}
\label{fig:efmfit}
\end{figure}

The vacuum correlator serves as a reference in this work.  To fix the
parameters of the lightest vector state in the finite volume of the
simulation, we fitted the vacuum correlation function to an
 Ansatz of the form
\be
G_{ii}(\tau,0)= A_1 e^{-m_1\tau} + \tfrac{3}{4\pi^2}\kappa_0\;\exp(-\Omega\tau)\;
\left( \Omega^2/\tau + 2\Omega/\tau^2 + 2/\tau^3\right),
\label{eqn:efmfit}
\ee
which is the Laplace transform of
% corresponding to the `one-state + continuum' form
\be
\frac{\rho_{ii}(\omega,0)}{2\pi} = A_1\; \delta(\omega-m_1) 
+ \tfrac{3}{4\pi^2}\theta(\omega-\Omega)\; \kappa_0 \,\omega^2.
\ee
The finite-volume spectral function consists of Dirac $\delta$
functions. However, since many states contribute at short distances,
the continuum in Eq.~(\ref{eqn:efmfit}) should be interpreted as
resulting from the contributions of many energy eigenstates.  In
Fig.~\ref{fig:efmfit} we show the vacuum vector correlator
$G_{ii}(\tau,T=0)$ and the fit result.  The insert shows the
corresponding effective mass.  The fit (performed in the interval
$5\leq \tau/a \leq 64$), clearly provides a good description of the
data. For $\tau/a\gtrsim32$ the quality of the data deteriorates and
the signal is lost.

We will only use the result for the parameters $A_1$ and $m_1$ in the
following. The question arises of the relation of these parameters to
the infinite-volume spectral function. An answer is given
by Refs.\ \cite{Luscher:1991cf,Meyer:2011um}. Assuming phenomenologically
reasonable values of the $\rho$ coupling to the $\pi\pi$ channel, the
mass $m_1$ in fact appears to be within $10\%$ of the
(infinite-volume) $\rho$ mass~\cite{Luscher:1991cf}.

In addition we estimate the thermal (isovector) quark number
susceptibility $\chi_s/T^2$ from the time-time component of the vector
correlation function (see Eq.~\ref{eq:chis}), by reading off the
midpoint of the time-time correlator. We checked the obtained value by
fitting the correlator to a constant in the region $2\leq\tau/a\leq14$
and found only a negligible deviation.  The parameters used in our
lattice setup, the `$\rho$-meson' parameters and the value of the
static susceptibility are summarized in Tab.~\ref{tab:pars}.

\begin{table}[t]
\centering
 % Give a unique label
% For LaTeX tables use
\begin{tabular}{|c|r|r|c|}
\hline
   &   $64^3\times 128$  & $64^3\times 16$  & Ref. \\
\hline
$6/g_0^2$ & \multicolumn{2}{|c|}{ 5.50} & \\
$\kappa$ & \multicolumn{2}{|c|}{ 0.13671 } & \\
$c_{\rm sw}$ & \multicolumn{2}{|c|}{ 1.7515} & \\ 
$Z_V$    & \multicolumn{2}{|c|}{  0.768(5)} & \cite{DellaMorte:2005rd}\\
$a[\textrm{fm}]$ & \multicolumn{2}{|c|}{ 0.0486(4)(5)} &   \cite{Fritzsch:2012wq}\\
$m_\pi$[MeV]   &  \multicolumn{2}{|c|}{270}   & \cite{Fritzsch:2012wq} \\
\hline
$T \; [{\rm MeV}]$ &  & 253(4) &\\ 
$\chi_s/T^2$ &  &  0.871(1) & \\
$A_1/T^3$ &  4.42(31) & &\\
$m_1/T$ &  3.33(5) & &\\
$\kappa_0$    & 1.244(5)& & \\
$\Omega/T$ & 5.98(11) & &\\
\hline
\end{tabular}
\caption{\it{% $N_{conf}[N_\tau=16]$ & 317 & \\
The top part of the table shows the common  quantities characterizing 
 the zero-temperature and finite-temperature ensembles.  In the lower part,
the fit parameters for the vacuum correlator in units of $T=253{\rm
  MeV}$ and the value of the (isospin) quark number susceptibility
$\chi_s/T^2$ are given.  For more details on the generation of the $N_\tau=128$ ensemble,
see \cite{Fritzsch:2012wq}.  The number of configurations generated
with $N_\tau=16$ is 317. }}
\label{tab:pars}      
\end{table}

%%%%%%%%%%%%%%%%%%%%%%%%%%%%%%%%%%%%%%%%%%%%%%%%%%%%%%%%%%%%%%%%%%%%%%%%
%%%%%%%%%%%%%%%%%%%%%%%%%%%%%%%%%%%%%%%%%%%%%%%%%%%%%%%%%%%%%%%%%%%%%%%%
%%%%%%%%%%%%%%%%%%%%%%%%%%%%%%%%%%%%%%%%%%%%%%%%%%%%%%%%%%%%%%%%%%%%%%%%

\subsection{Thermal and vacuum correlators}
\label{sec:data}

\begin{figure}[t]
\centering
\includegraphics[width=.63\textwidth]{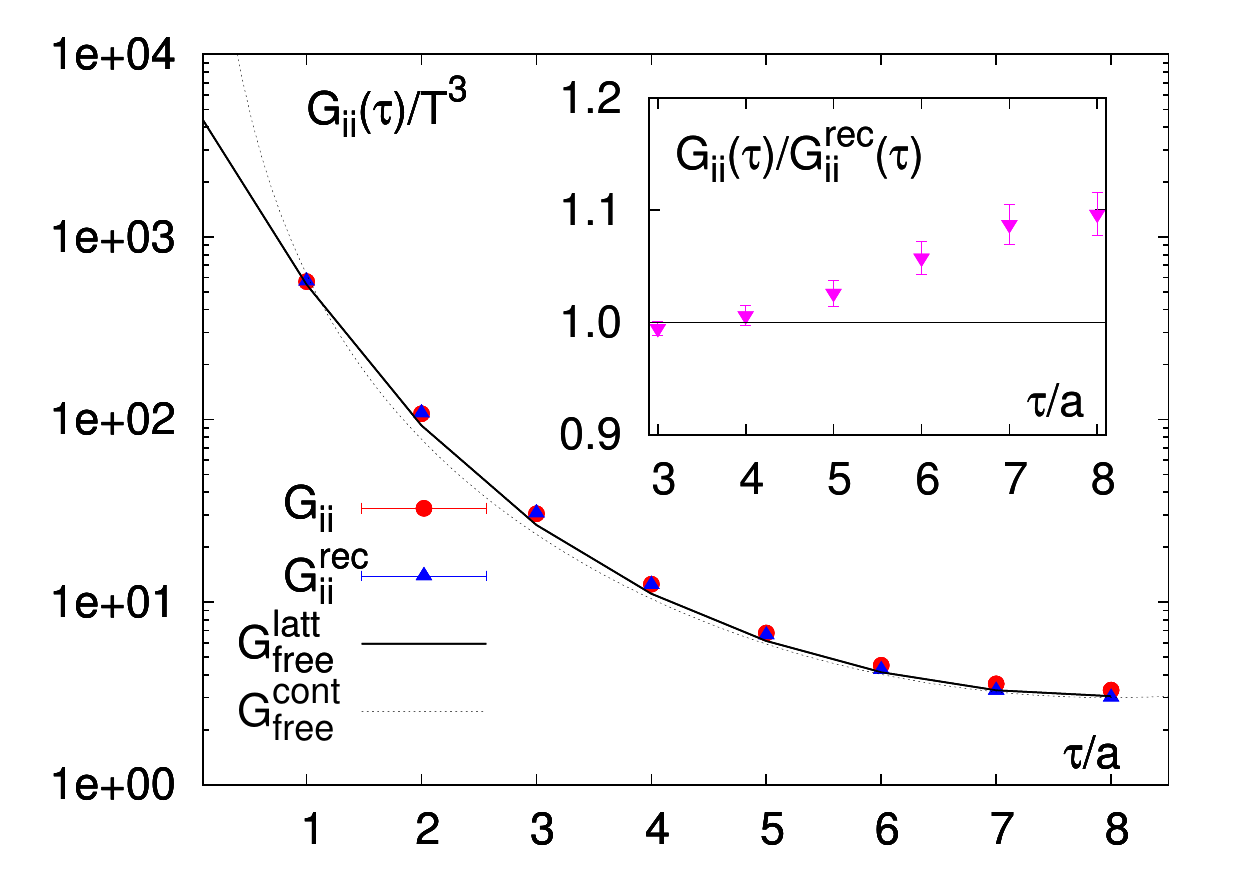}
\caption{\it{Thermal $G_{ii}(\tau)/T^3$ and reconstructed
    $G_{ii}^{rec}(\tau)/T^3$ vector correlators over Euclidean time
    separation $\tau$ compared to the free (continuum and lattice)
    cases. The reconstructed correlator was computed by applying 
    Eq.~\ref{eq:Grec1-main} to the data obtained from a lattice sized
    $N_\sigma=64$ and $N_\tau=128$ . The insertion shows the ratio
    $G_{ii}(\tau)/G_{ii}^{rec}(\tau)$.}}
\label{fig:Correlators}
\end{figure}

In Fig.~\ref{fig:Correlators} we show the correlator $G_{ii}(\tau,T)$
computed at $T\simeq250$MeV together with the corresponding free
`continuum' and free `lattice discretized' correlation functions. In
addition we show the reconstructed correlator $G_{ii}^{rec}(\tau)$ as
obtained from Eq.~\ref{eq:Grec2-main}. The results for the thermal and
reconstructed correlators for $\tau/a\geq4$ can be found in
Tab.~\ref{tab:results}.  

The reconstructed correlator lies somewhat lower than the thermal
correlator.  The insert in Fig.~\ref{fig:Correlators} displays the
ratio $G_{ii}(\tau)/G_{ii}^{rec}(\tau)$ in order to make their
relative $\tau$ dependence visible. For small Euclidean times
$\tau<\beta/4$ this ratio is unity, above it increases monotonically
until it levels off around the midpoint at about $10\%$ above unity.
A thermal modification of the spectral function has thus taken place
(recall that the spectral function underlying $G_{ii}^{rec}(\tau)$
contains the bound states of the confined theory).

\begin{figure}[t]
\centering
\includegraphics[width=.63\textwidth]{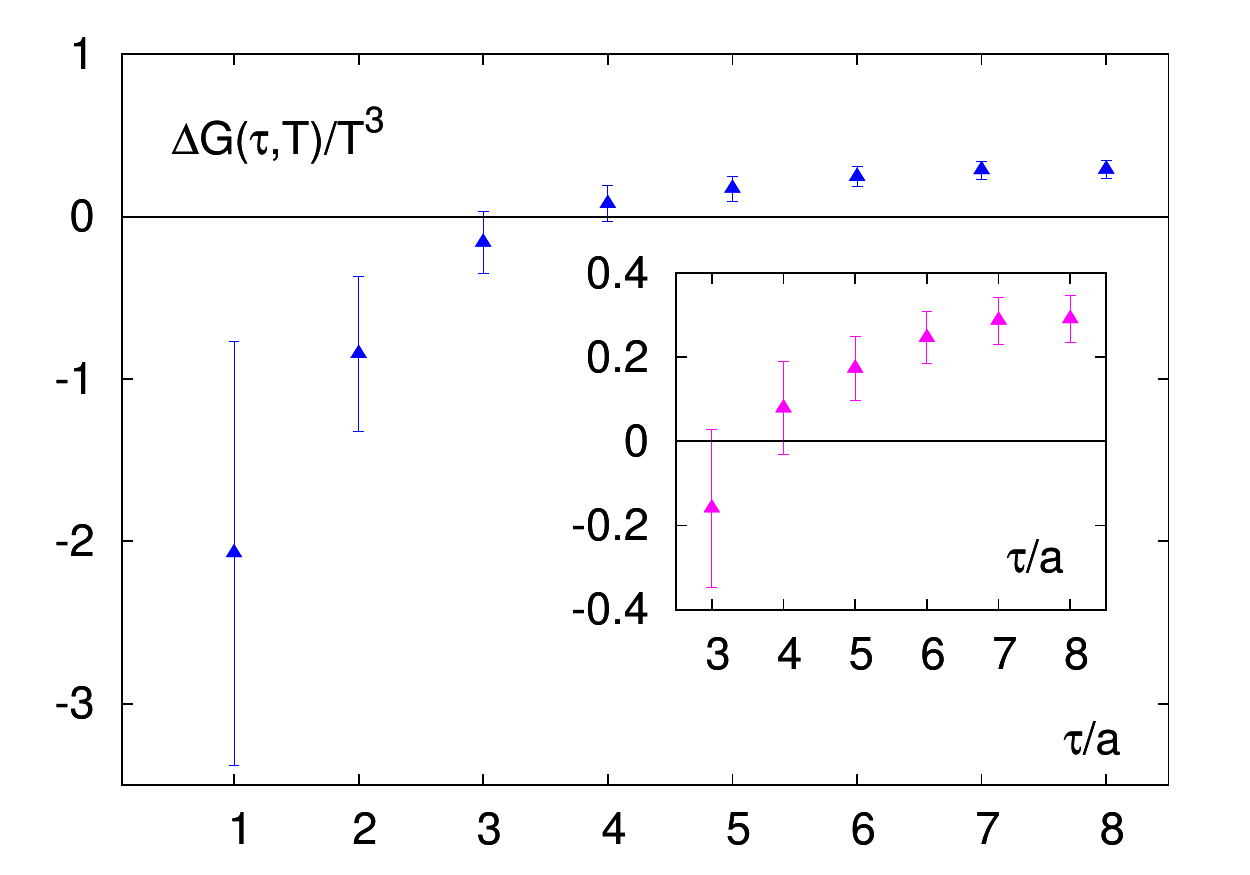}
\caption{\it{Difference   $\Delta G(\tau,T)/T^3$  
  of the thermal vector correlator and the
  corresponding reconstructed correlator
  as a function of Euclidean time
  $\tau$. The insertion shows $\Delta G(\tau,T)/T^3$
  in the region $\tau/a\ge 3$. }}
\label{fig:GdiffGrec}
\end{figure}

\begin{table}[t]
\centering
 % Give a unique label
% For LaTeX tables use
\begin{tabular}{lll}
\hline\noalign{\smallskip}
$\tau$ & $G_{ii}(\tau)/T^3$ & $G_{ii}^{rec}(\tau)/T^3$ \\
\noalign{\smallskip}\hline\noalign{\smallskip}
4 & 12.534(94) & 12.456(59) \\
5 & 6.775(65) & 6.602(41) \\
6 & 4.510(51) & 4.264(35) \\
7 & 3.569(45) & 3.282(33) \\
8 & 3.300(44) & 3.008(34) \\
\noalign{\smallskip}\hline
\end{tabular}
\caption{\it{Results of the thermal and reconstructed correlation functions
 for $\tau/a\geq4$. Note that all results have been renormalized using the value
 of $Z_V$ in Table (\ref{tab:pars}) and normalized by $T^3$.}}
\label{tab:results}      
\end{table}

In Fig.~\ref{fig:GdiffGrec} we show the difference 
\be
\Delta G(\tau,T) \equiv 
G_{ii}(\tau,T)-G_{ii}^{rec}(\tau,T)=\int_0^\infty \frac{d\omega}{2\pi} 
\Delta\rho(\omega,T)
\frac{\cosh[\omega(\beta/2-\tau)]}{\sinh(\omega\beta/2)}
\label{eq:GdiffGrec}
\ee
of the thermal and the reconstructed correlators. Given that
$\rho_{ii}(\omega,T)$ and $\rho_{ii}(\omega,T=0)$ have the same
$\sim\omega^2$ behavior, this means we are subtracting
non-perturbatively the ultraviolet tail of the spectral
function. Using this difference we are therefore able to probe
the change in the vector spectral function from the confined
to the deconfined phase for frequencies $\omega \lesssim {\rm O}(T)$.

For small times, the difference (\ref{eq:GdiffGrec}) turns out to be negative,
while it is positive for $\tau\geq\beta/4$. Note the
errors decrease with increasing Euclidean time throughout the
available range.  We show a more detailed view of the region $\tau/a\geq 3$
in the insert of Fig.~\ref{fig:GdiffGrec}. Here the difference
still exhibits a mild increase and levels off near the midpoint.
The value it reaches at the midpoint is
$\Delta G(\tau=\beta/2,T)/T^3=0.291(55)$.

\begin{figure}[t]
\centering
\hspace*{-0.8cm}
\includegraphics[width=.53\textwidth]{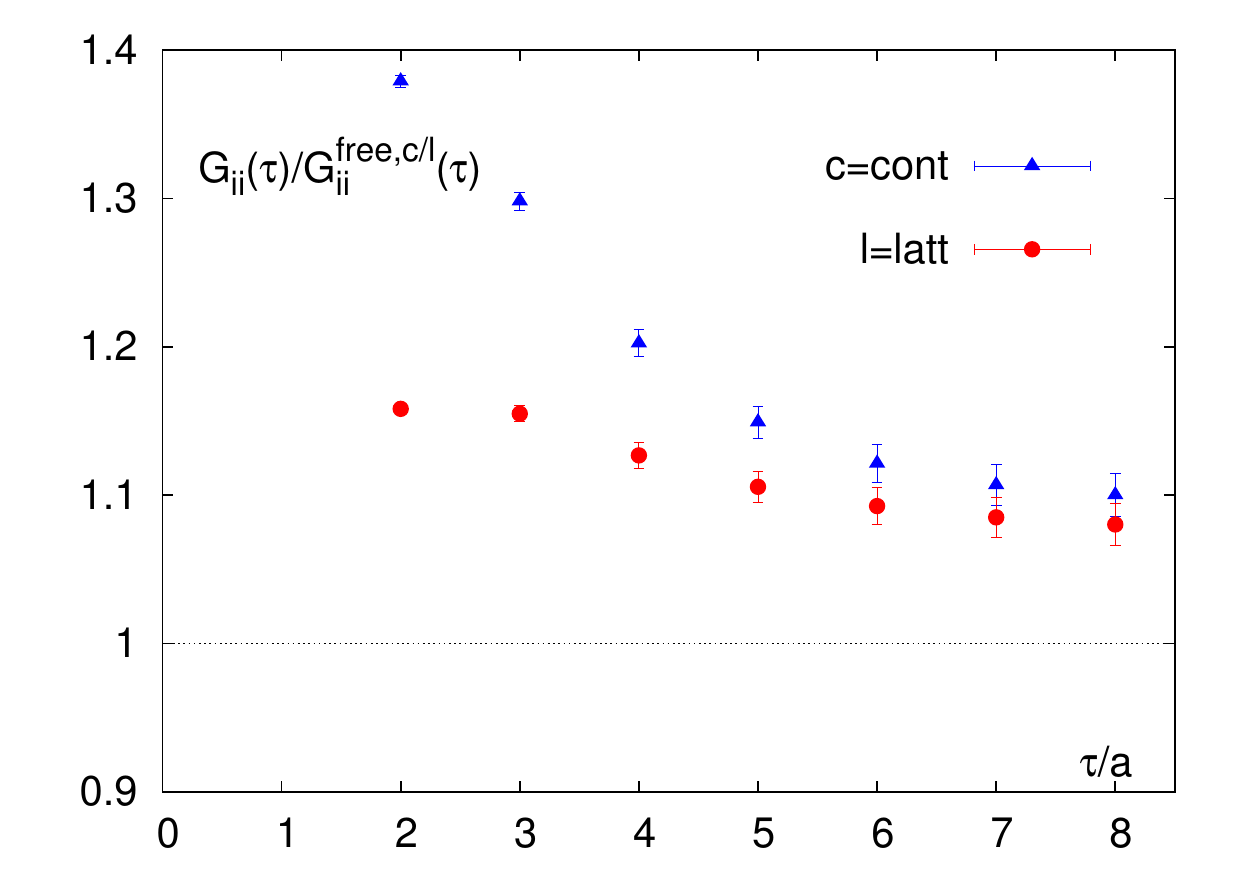}
\hspace*{-0.8cm}
\includegraphics[width=.53\textwidth]{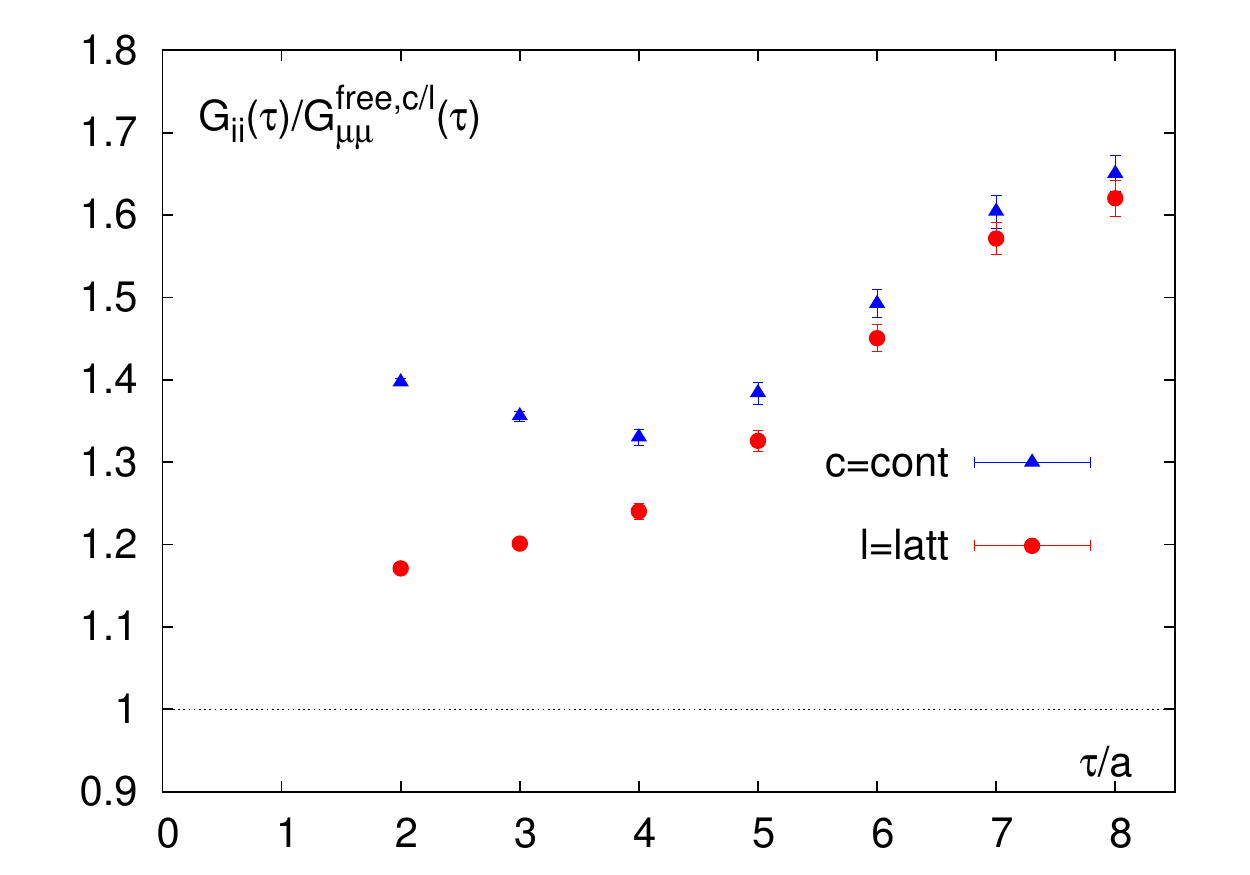}
\caption{\it{Left: The vector correlator $G_{ii}(\tau)$ at $T\simeq250$MeV 
normalized by the free continuum and free discretized correlation functions
 $G_{ii}^{free,c/l}$. Right: The vector correlator $G_{ii}(\tau)$ 
normalized by the free (continuum/discretized) correlation 
functions $G_{\mu\mu}^{free,c/l}$.}} 
\label{fig:GbyGfree}
\end{figure}

\subsection{Comparison with the free-quark correlator}

In the previous subsection we compared the thermal correlator to
its zero-temperature analogue.  Now we analyze the thermal correlator
in relation to the non-interacting case, which by asymptotic freedom
corresponds to the regime of asymptotically high temperature.
However, before discussing the departure of the simulation data from
the non-interacting case, we address briefly the issue of cutoff
effects. 

Since we only have data at one lattice spacing, the value of the
lattice correlator, viewed as an estimator of the continuum
correlator, is necessarily ambiguous at some level due to cutoff
effects.  To estimate the size of this ambiguity, we show the ratio
$G_{ii}/G_{ii}^{free,c/l}$ in Fig.~\ref{fig:GbyGfree}(left), where
$c$ and $l$ denote the analytically known free continuum and free lattice
cases, respectively \cite{Aarts:2005hg}.  Concentrating on the ratio to
the free \emph{continuum} correlator we observe a decreasing trend
throughout the entire available Euclidean time range, a behavior very
similar to that seen in quenched studies \cite{Ding:2010ga}.  Taking
the ratio to the free \emph{lattice} correlation function on the other
hand the results are much flatter and almost constant at small times
$\tau/a\le3$, while for $\tau/a\ge5$ the two ratios track each other
and are separated by a small shift of roughly $2\%$.  The difference
between the two curves comes from the fact that the free lattice
correlation function takes into account the tree-level lattice
artifacts.

To put this systematic uncertainty into perspective, we examine the
ratio $G_{ii}/G_{\mu\mu}^{free,c/l}$ in the right panel of
Fig.~\ref{fig:GbyGfree}. The only difference between
$G_{ii}^{free,c/l}$ and $G_{\mu\mu}^{free,c/l}$ is that the
$\delta$-function in the spectral function at $\omega=0$ is absent in
the latter. The difference between the left and right panel curves
thus corresponds to the contribution of the transport peak.  It is
clear from the figure that this contribution is much larger than the
cutoff effects present at tree-level for $\tau/a\geq 5$.

Returning now to the question of how much the thermal correlator
differs from its non-interacting counterpart, we see that the
simulation data lies 8 to 10$\%$ above the free lattice correlator.
The sign of the effect corroborates our finding in section \ref{sec:phenorho}
that spectral weight is missing from the weak coupling spectral function.

% Since the leading radiative correction to the spectral function in the
% ultraviolet is a factor $(1+\alpha_s/\pi)$, it is possible that
% next-to-leading order perturbation theory can describe the simulation data.

\subsection{Thermal moments of the correlator\la{sec:moments}}

Finally we compute also the ratio of thermal moments of the correlator \cite{Ding:2010ga}:
\be
R^{(2,0)}=\frac{G^{(2)}}{G^{(0)}}=2\frac{\int d\omega \rho(\omega)\overline{K}(\omega,T)}
{\int d\omega \omega^2\rho(\omega)\overline{K}(\omega,T)}\quad
\textrm{where}\quad \overline{K}(\omega,T)=\frac{1}{2\pi\sinh(\omega/2T)}~~~.
\ee
This quantity can be extracted from the correlator data by combining the results 
of $(G_{ii}/G_{ii}^{free,c/l})(\tau=\beta/2)=G_{ii}^{(0)}/G_{ii}^{(0),free}$ and:
\be\la{eq:squ}
\Delta_{ii}(\tau) \equiv   
\frac{G_{ii}(\tau)-G_{ii}^{(0)}}
{G_{ii}^{free}(\tau )-G_{ii}^{(0),free}}  
=
\frac{G^{(2)}_{ii}}{G^{(2),free}_{ii}}
\left[ 1+{\rm O}((\beta/2-\tau)^2) \right].
% \left( 1+ \left( R^{(4,2)} -R^{(4,2)}_{free} \right)
% \left(\frac{1}{2T} - \tau \right)^{2} + ...\right) \; .
\ee
In the free case a straightforward computation \cite{Florkowski:1993bq} 
yields $R^{(2,0)}_{free}=18.423$. Using this result together with our lattice 
data we obtain:
\be
G_{ii}^{(0)}/G_{ii}^{(0),free}=1.100(15) ~,~~~ G_{ii}^{(2)}/G_{ii}^{(2),free}=
1.198(8)\quad\Rightarrow \frac{R^{(2,0)}}{R^{(2,0)}_{free}}=1.089(30)~~,
\ee
where we neglected higher-order terms in the square bracket of
Eq.\ (\ref{eq:squ})\footnote{Including the leading correction into a fit we
  find it to be poorly determined by the data, while the constant
  contribution remained unchanged within errors.}.  We see that the
ratio $R^{(2,0)}$ of thermal moments is roughly $6-12\%$ larger than
the free result. As the ratio of second thermal moments is sensitive
to changes in the low frequency region of the spectral function (see
e.g. \cite{Ding:2010ga}), this observation could be due to a
broadening of the $\delta$-function form of the free theory.

%%%%%%%%%%%%%%%%%%%%%%%%%%%%%%%%%%%%%%%%%%%%%%%%%%%%%%%%%%%%%%%%%%%%%%%%
%%%%%%%%%%%%%%%%%%%%%%%%%%%%%%%%%%%%%%%%%%%%%%%%%%%%%%%%%%%%%%%%%%%%%%%%
%%%%%%%%%%%%%%%%%%%%%%%%%%%%%%%%%%%%%%%%%%%%%%%%%%%%%%%%%%%%%%%%%%%%%%%%
\section{Analysis of lattice correlators in terms of spectral functions}

We begin with a simple but instructive analysis of the spectral
function difference $\Delta\rho$. Section \ref{sec:GGrec} contains the
main analysis based on fits to the thermal part of the Euclidean
correlator, and section \ref{sec:fitG} describes a fit directly to the
thermal correlator. The results of the two fits are compared against
each other and against previous quenched calculations in section
\ref{sec:disc}.

\subsection{A simple spectral analysis of the thermal part of the vector correlator\la{sec:simp}}

\begin{figure}[t]
\centerline{\includegraphics[width=.63\textwidth]{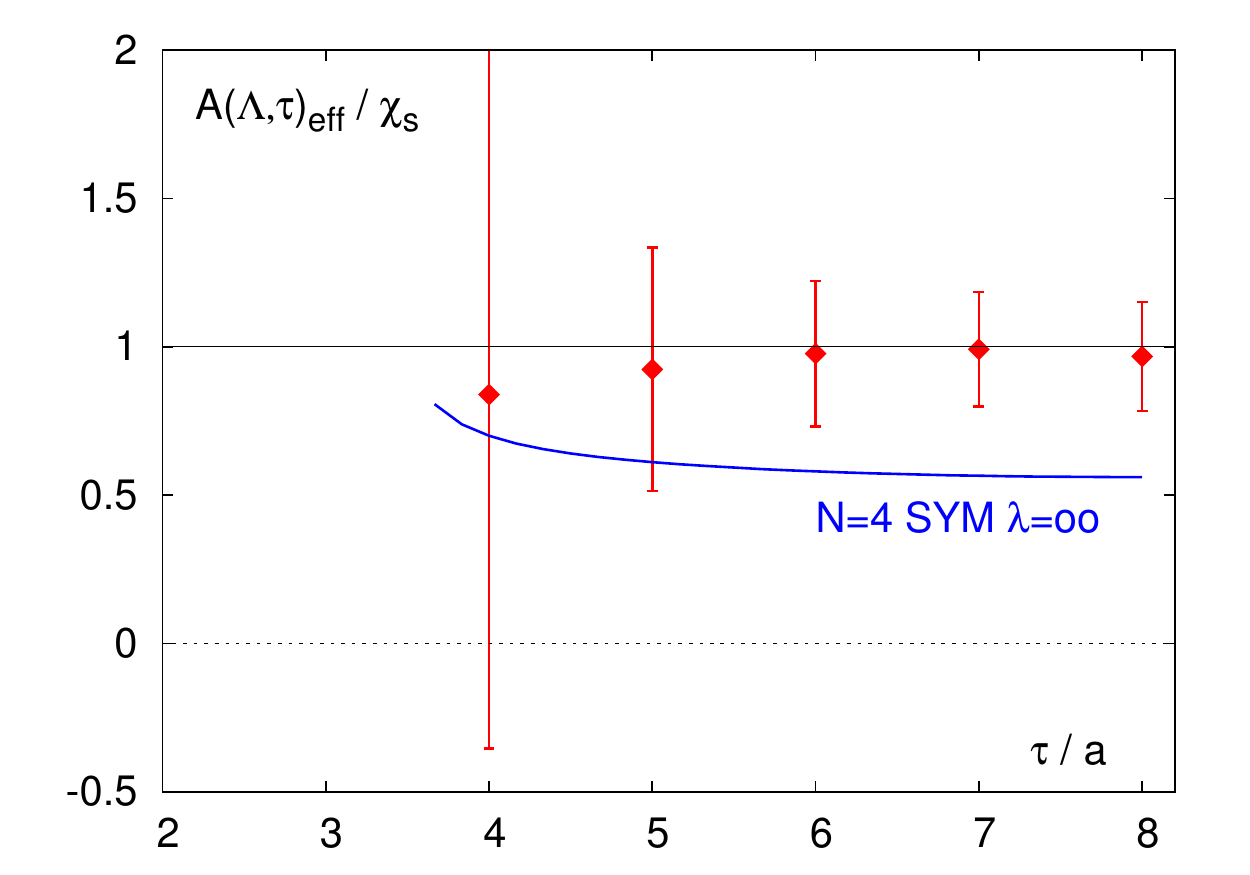}}
\caption{\it\small{The effective area ${\cal A}(\Lambda,\tau)_{\rm eff}$,
defined in Eq.\ (\ref{eq:DGsimp2}), normalized by the static susceptibility.
The corresponding quantity in the strongly coupled SYM theory calculated from
Eq.\ (\ref{eq:N4vc}) is also displayed for comparison.}} 
\label{fig:data-cosh}
\end{figure}

Based on the analysis of section \ref{sec:phenorho}, it is interesting
to ask whether the Euclidean correlator can be described and the sum
rule (\ref{eq:sr}) satisfied solely by the transport peak and the
$\rho$-meson contribution. For this purpose, we consider the following
caricature of $\Delta\rho(\omega,T)$, where the sum rule has already
been enforced, 
\be\la{eq:SFsimp} 
\frac{\Delta\rho(\omega,T)}{2\pi} =
C\, \omega \left(\delta(\omega) -
{\txts\frac{1}{2}}\delta(\omega-m_1)\right), 
\ee 
corresponding to the Euclidean correlator 
\be
\la{eq:DGsimp} \Delta G(\tau,T) = C \left(\frac{1}{\beta} - \frac{m_1}{2}
\frac{\cosh m_1(\beta/2-\tau)}{\sinh m_1\beta/2}\right).  
\ee 
The mass $m_1$ is set equal to the value obtained by fitting the
Ansatz (\ref{eqn:efmfit}) to the vacuum correlator and given in table
\ref{tab:pars}.  The kinetic theory sum rule (\ref{eq:IntLbda})
implies that $C$ is an estimator for ${\cal A}(\Lambda)$, and from
Eq.\ (\ref{eq:DGsimp}) an effective value ${\cal A}(\Lambda,\tau)_{\rm
  eff}$ can thus be defined for every value of $\tau$,
\be
\la{eq:DGsimp2}
{\cal A}(\Lambda,\tau)_{\rm  eff} = 
 \frac{\Delta G(\tau,T)}{\left(\frac{1}{\beta} - \frac{m_1}{2}
\frac{\cosh m_1(\beta/2-\tau)}{\sinh m_1\beta/2}\right)} .
\ee 
The result is shown in Fig.\ (\ref{fig:data-cosh}).  The quantity
${\cal A}(\Lambda,\tau)_{\rm eff}$ is well compatible with a constant
value, implying that the Ansatz (\ref{eq:SFsimp}) already provides a
good description of the lattice correlator. Moreover, the
weak-coupling expectation that ${\cal A}(\Lambda,\tau)_{\rm eff}$
should be given by $\chi_s$ (up to quark mass effects reducing their
average thermal velocity $\<v^2\>$) is also well reproduced.

It should be noted that in this simple picture the prefactor of 
$\delta(\omega-m_1)$  in Eq.\ (\ref{eq:SFsimp}) is about $-\chi_s m_1/2 =
-1.45(2)T^3$, to be compared with the area $A_1=+4.42(31)T^3$ obtained
from the vacuum correlator (table \ref{tab:pars}). This observation
means that the area under the thermal spectral function in the region
$T\lesssim \omega \lesssim 4T$ cannot be negligible compared to $A_1$,
confirming the conclusions drawn from phenomenology in section
(\ref{sec:phenorho}).

It is interesting to confront the lattice data with the spectral
function in the strongly coupled SYM theory, Eq.\ (\ref{eq:N4vc}).  We
therefore also plot the corresponding SYM function in
Fig.\ (\ref{fig:data-cosh}). It lies lower than the lattice data, in
spite of the fact that $\rho^{\rm R}_{ii}(\omega,T)/(\chi_s\omega^2)$
has the same large-frequency limit as $\rho_{ii}^{\rm
  free}(\omega,T)/(\chi^{\rm free}_s\omega^2)$ in QCD and also
satisfies the sum rule (\ref{eq:sr}). This comparison shows that the
lattice data is not simultaneously compatible with the combination of
(a) the functional form of the SYM spectral function and (b) the
corresponding very low diffusion constant. The data is however
perfectly compatible with the substitution of the delta function in
Eq.\ (\ref{eq:SFsimp}) by a flat (or even $\propto
(1+\frac{1}{24}\omega^2/T^2)$ as in the SYM case) behavior of
$\Delta\rho(\omega,T)/\omega$ up to about $\omega\approx 4T$, provided
its area is adjusted appropriately.

In order to parametrize $\Delta\rho(\omega,T)$ systematically,
including the expected contributions at high frequencies, we resort to
the more sophisticated fits described in the next subsection. To
anticipate the results, similar qualitative conclusions on the
distribution of the spectral weight in $\Delta\rho(\omega,T)$ will be
obtained.

\subsection{Fit to the thermal part of the vector correlator}
\label{sec:GGrec}

We proceed to investigate the behavior of the thermal part of the
spectral function $\Delta\rho$ by fitting the difference of the
thermal and the reconstructed correlator, see
Eq.\ (\ref{eq:GdiffGrec}). We know from the operator-product expansion
that the difference of spectral functions falls off rapidly (as
$\omega^{-2}$) for $\omega\gg T$, and therefore focus on the region
$\omega \lesssim {\rm O}(T)$ in order to choose a fit Ansatz.  As described in the previous
subsection, the fact that the data (displayed in
Fig.~\ref{fig:GdiffGrec}) is positive at long distances and negative
at short distances suggests that the thermal spectral weight exceeds
the vacuum spectral weight at low frequencies and falls short of it at
higher frequencies\footnote{A continuum extrapolation is really needed
  to confirm the behavior at short distances.}.

We thus parametrize $\Delta\rho$ using the following Ansatz for $\omega\geq 0$:
\ba \la{eq:ansatz2}
 \Delta\rho(\omega,T)&=&\rho_T(\omega,T)-\rho_B(\omega,T)+\Delta\rho_F(\omega,T), 
 \phantom{\frac{99}{17}}\\  \la{eq:ansatz3}
{\rho_B(\omega,T)} &=& \frac{2 c_B\, g_B\,\tanh(\omega/T)^3}{ 4(\omega-m_B)^2 + g_B^2 },
 \\   \la{eq:ansatz4}
 {\rho_{T,1}(\omega,T)} &=& \frac{4 c\,\omega}{ (\omega/g)^2 + 1 },\qquad \qquad ~~~~~\,
 {\rho_{T,2}(\omega,T)}=\frac{4c\,T\tanh(\omega/T)}{ (\omega/g)^2 + 1 },
\\    \la{eq:ansatz5}
\Delta\rho_F(\omega,T) &=& \rho_F(\omega,T) - \rho_F(\omega,0),\qquad 
{\rho_F(\omega,T)} = \frac{3}{2\pi}\kappa\,\omega^2 \tanh\left(\frac{\omega}{4T}\right)\;.
\ea
The bound state (B) and the transport peak (T) are represented by
Breit-Wigner forms. Even such a simple Ansatz requires three
parameters $(c_B,g_B,m_B)$ to determine the bound state peak, two
parameters $(c,g)$ for the transport peak and one $(\kappa)$ for the `perturbative'
contribution (F). We will therefore fix some of them using the vacuum
correlator.  In the following we set $m_B$ equal to $m_1$, given in
Tab.~\ref{tab:pars}, which we obtained from the exponential fit to the
vacuum correlator.  Note that the area under the bound state peak
$\int dw~\rho_B/\omega$ does not depend on the width $g_B$ in the
limit where it is small.  The sensitivity of the Euclidean correlator
to the latter parameter is very small. We therefore perform fits for
three fixed values of this parameter, and check the sensitivity of the
result.  We choose the values $g_B/T=0.1, 0.5$ and 1.0, corresponding
to $g_B\simeq 25, 125$ and 250MeV.

\begin{figure}[t]
\centerline{
\includegraphics[width=.53\textwidth]{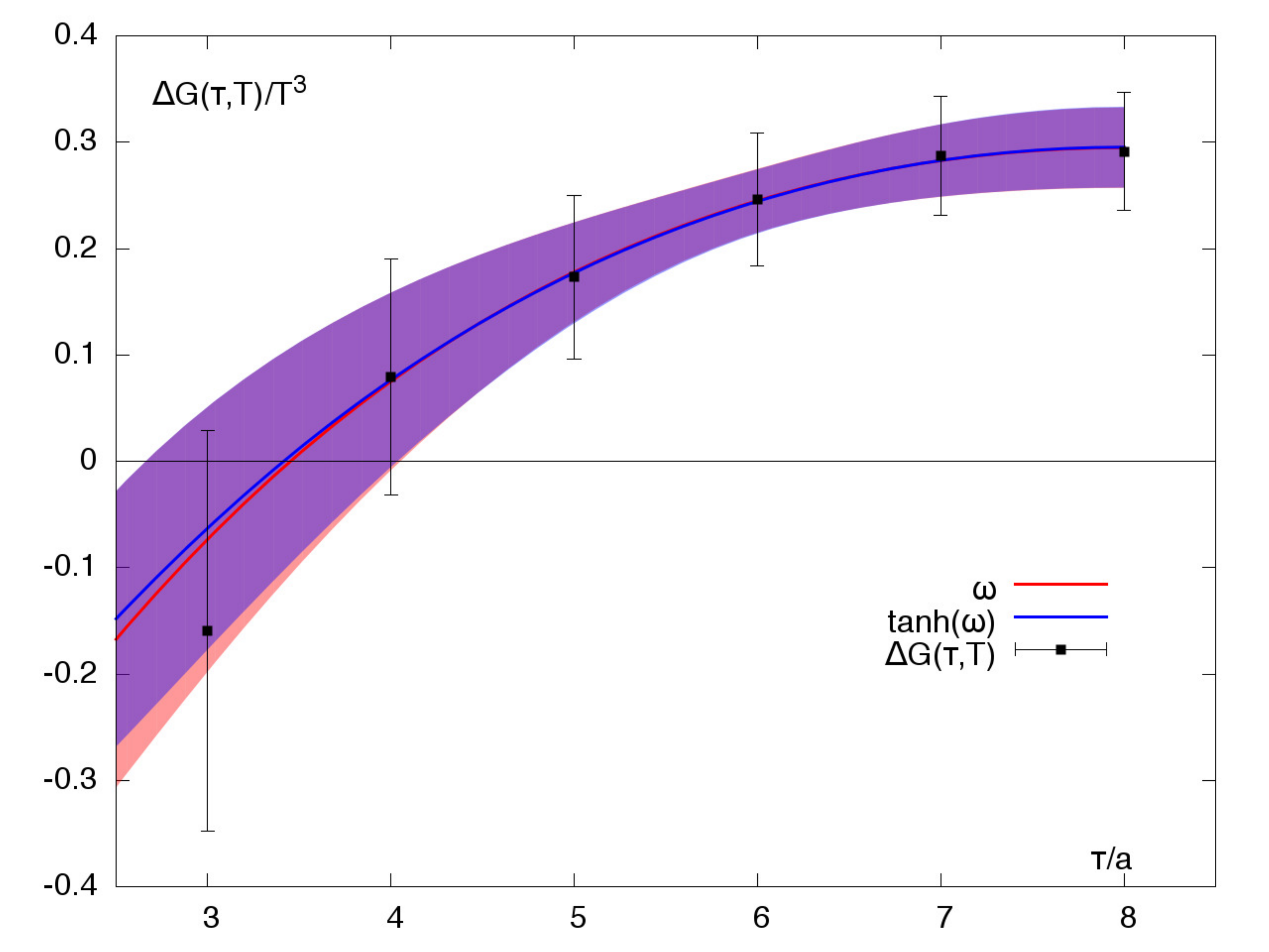}
\hspace*{-0.5cm}
\includegraphics[width=.53\textwidth]{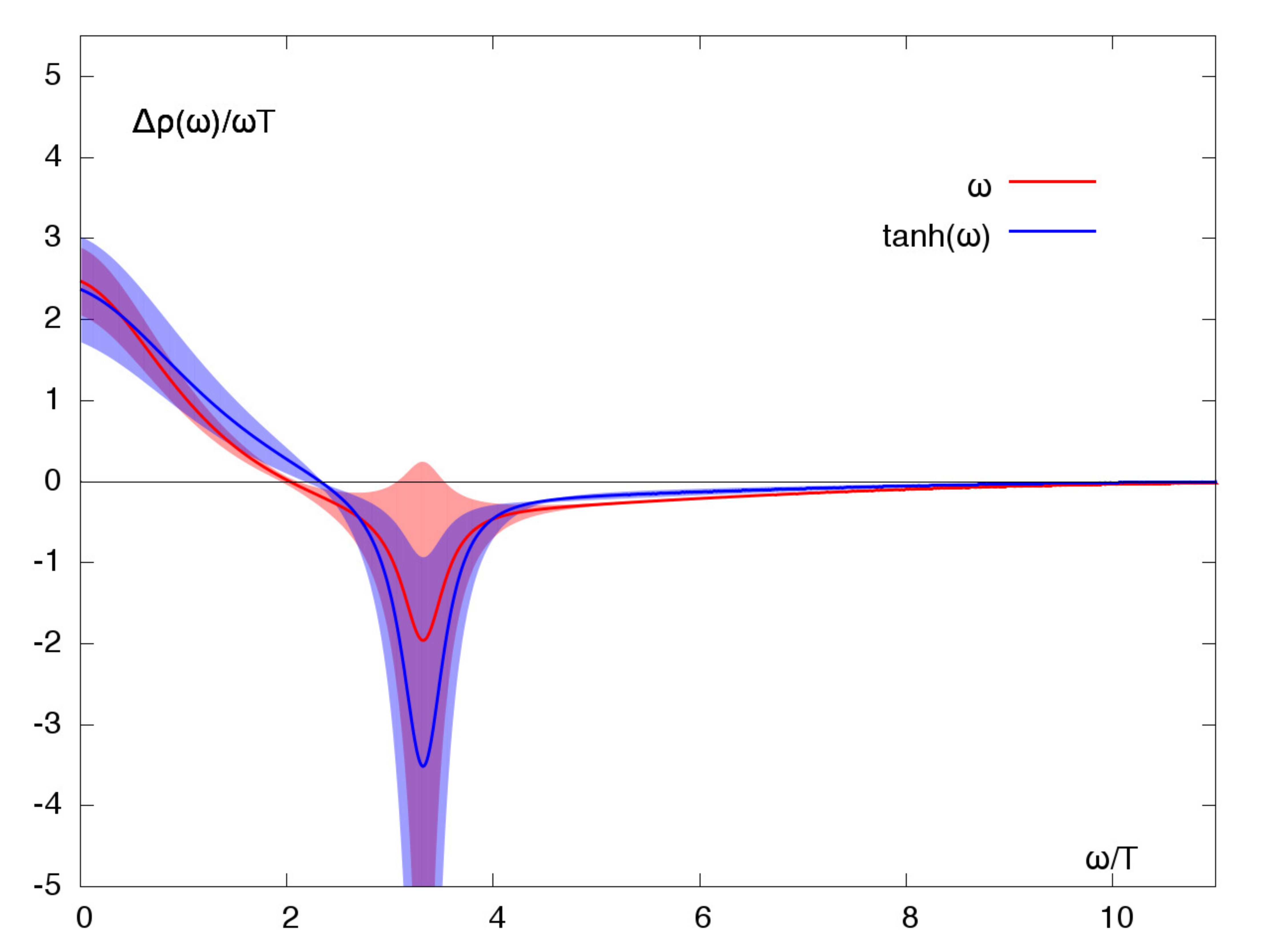}}
\caption{\it\small{Left: Fits to $\Delta G(\tau,T)/T^3\equiv
    [G_{ii}(\tau)-G_{ii}^{rec}(\tau)]/T^3$. The blue and red results
    differ by the form of transport peak in the Ansatz. The error
    bands are computed from the covariance matrix of the fit.  Right:
    The resulting spectral functions for both Ans\"atze.}}
\label{fig:GdiffGrec-CORSPF}
\end{figure}

The tail $\sim T/\omega$ of the Ansatz $\rho_{T,1}$ violates
the OPE prediction that $\Delta\rho\sim(T/\omega)^2$ at large
frequencies.  It has been argued in \cite{Burnier:2012ts} that this
might lead to an overestimate of the transport contribution. To avoid
this problem we introduce the Ansatz 2, where $\omega\rightarrow
T\tanh(\omega/T)$. This Ansatz possesses the correct asymptotic
behavior, as well as the expected linear behavior in $\omega$ at small
frequencies.

Finally, to complete the parametrization of $\Delta\rho(\omega)$, we
include a weak-coupling term inspired by Eq.~(\ref{eq:SFfree})
describing the subtraction of the large frequency parts of the thermal
and vacuum spectral functions. This contribution
$\rho_F(\omega,\kappa)\rightarrow 0$ vanishes exponentially as the
frequency increases.

In the next step we fit the combined Ans\"atze of
$\Delta\rho(\omega,c_B,g_B,m_B,c,g,\kappa)$ to the data, while at the
same time satisfying the sum rule of Eq.~\ref{eq:sr} to an accuracy of
$10^{-8}$.  We limit ourselves to fitting the region $5\leq\tau/a\leq8$
only, in order to minimize the influence of cut-off effects, as
discussed in Sec.~\ref{sec:data}.  With $m_B$ determined by the vacuum
correlator, we set $g_B$ successively to the three different values
mentioned above and fixed  $\kappa$  around unity, and fitted
the parameters $c$, $g$ and $c_B$.  The errors and error bands shown
in the following have been computed using the covariance matrix of the
corresponding fit for fixed values of $g_B$ and $\kappa$.

The resulting correlators and spectral functions are displayed in
Fig.~\ref{fig:GdiffGrec-CORSPF} for $g_B/T=0.50$ and $\kappa=1.10$
while the fitted parameters are given in Tab.~\ref{tab:fits}. In the
left panel of Fig.~\ref{fig:GdiffGrec-CORSPF} the data $\Delta
G(\tau,T)$ is compared to the fits using $\rho_{T,1}(\omega)$ and
$\rho_{T,2}(\omega)$ as transport contribution. We achieve a
quasi-perfect description of the data for $\tau/a\geq 4$.  The right
panel shows that both Ans\"atze exhibit a substantial spectral weight
around the origin and a negative contribution from the region of the
$\rho$ mass.

\begin{figure}[t]
\centerline{\includegraphics[width=.53\textwidth]{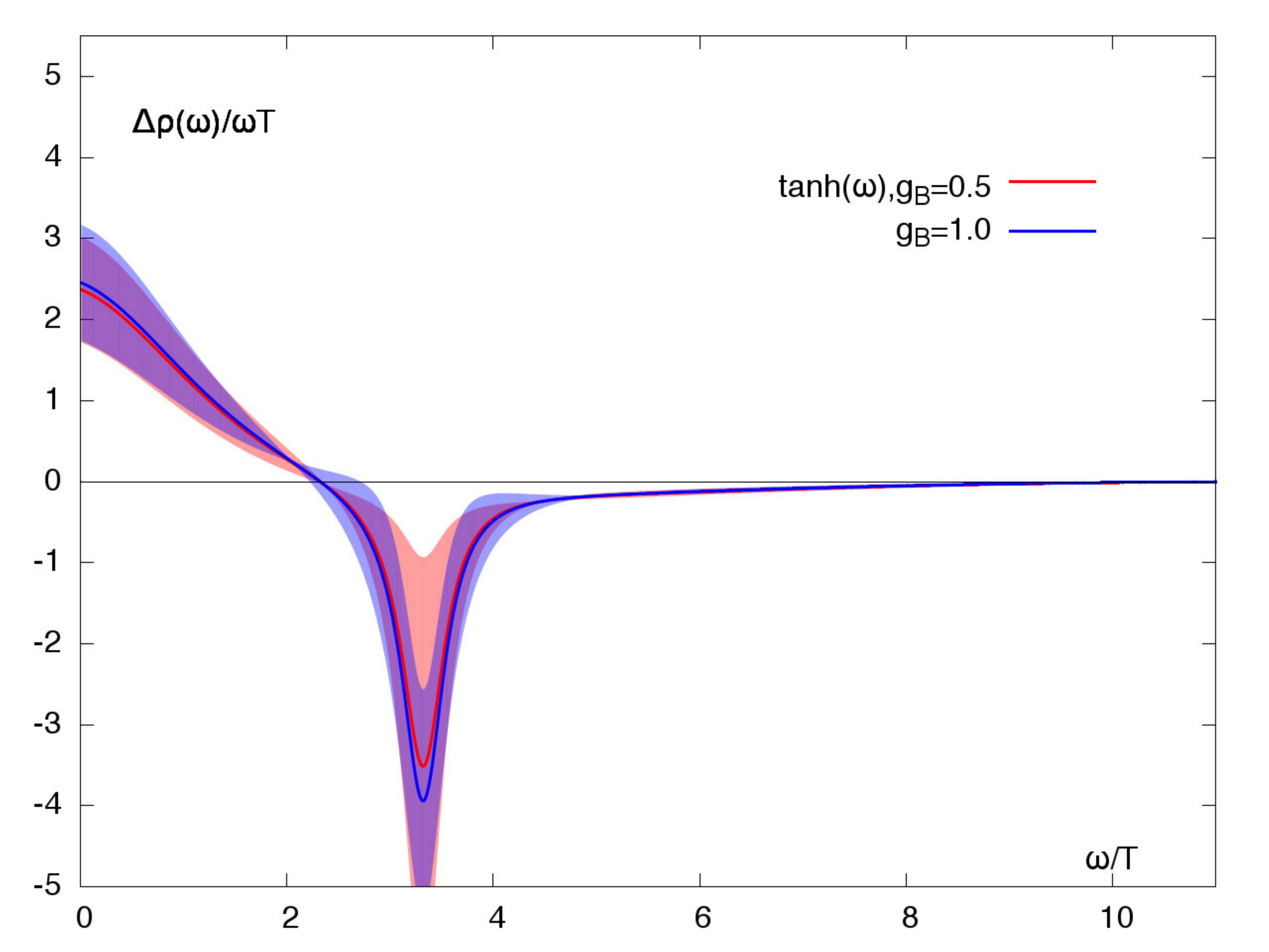}
\hspace*{-0.5cm}
\includegraphics[width=.53\textwidth]{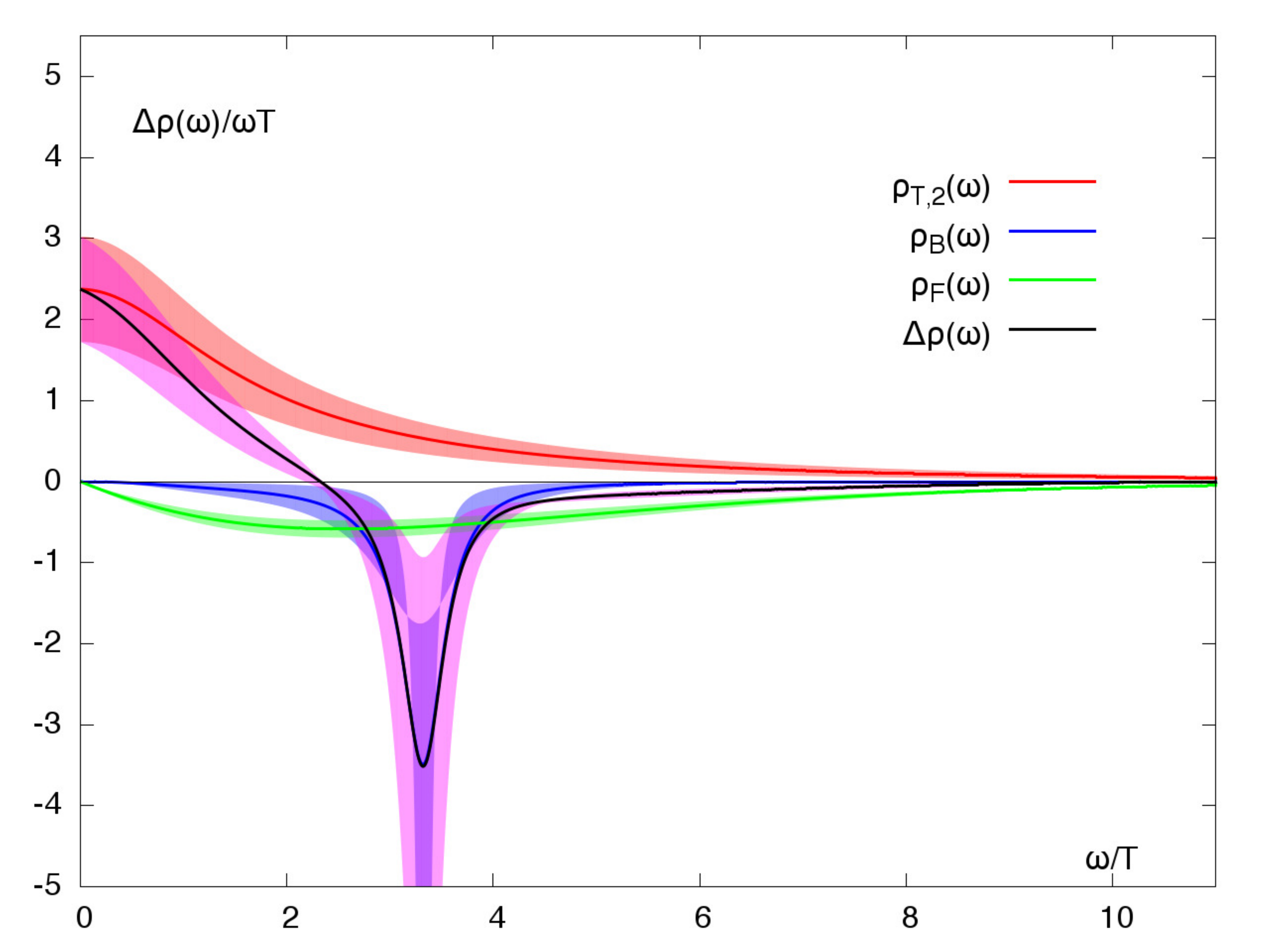}}
\caption{\it\small{Left panel: the impact of changing the width of the 
peak representing the $\rho$ resonance, $g_B=0.5T$ and $g_B=1.0T$.
Right panel: the separate contributions
    $\rho_T(\omega),~\rho_B(\omega),~\rho_F(\omega)$ and the full
    result of the fitted spectral functions for the transport
    Ansatz $\rho_{T,2}$.}}
\label{fig:GdiffGrec-SPF}
\end{figure}

Focussing on Ansatz 2, the left panel of Fig.~\ref{fig:GdiffGrec-SPF}
shows the sensitivity of the fit result to varying the parameter
$g_B$.  Varying the width only has a small effect on the overall
result. We also found little sensitivity to $\pm 15\%$ variations in
$\kappa$.  In order to understand what drives the parameters to their
final fitted values, we also plot separately the three contributions
appearing in Eq.\ (\ref{eq:ansatz2}--\ref{eq:ansatz5}), including
their respective error bands, in the right panel of
Fig.~\ref{fig:GdiffGrec-SPF}.  The contribution $\rho_F(\omega)$
mainly affects the intermediate frequency region around
$\omega/T\simeq2$. Its tail between $4\leq \omega/T\leq 10$ is largely
compensated by the tail of the Lorentzian centered at the origin, and
this might well be what drives the width of the Lorentzian.

%%%%%%%%%%%%%%%%%%%%%%%%%%%%%%%%%%%%%%%%%%%%%%%%%%%%%%%%%%%%%%%%%%%%%%%%
%%%%%%%%%%%%%%%%%%%%%%%%%%%%%%%%%%%%%%%%%%%%%%%%%%%%%%%%%%%%%%%%%%%%%%%%
%%%%%%%%%%%%%%%%%%%%%%%%%%%%%%%%%%%%%%%%%%%%%%%%%%%%%%%%%%%%%%%%%%%%%%%%
\subsection{Weak-coupling inspired fit to the thermal vector\la{sec:fitG}}

In contrast to the previous section, here we study directly the
thermal vector correlator and its ratio of thermal moments
$R^{(2,0)}$.  We perform a fit inspired by the weak coupling form of
the thermal spectral function,
\ba
\rho(\omega,T)&=&\rho_T(\omega,T)+\rho_F(\omega,T)~,
\ea 
where the form of the two contributions is defined in
Eq.\ (\ref{eq:ansatz4}) and (\ref{eq:ansatz5}).  At a given temperature this Ansatz is
characterized by three parameters $(c,g,\kappa)$.  We fit the full
Ansatz $\rho(\omega,c,g,\kappa)$ to the thermal correlator
$G_{ii}(\tau)$, while at the same time demanding that $R^{(2,0)}$ be
reproduced. In this analysis the three parameters $c, g$ and $\kappa$
are fitted, and the fit range is $5\leq\tau/a\leq 8$ as before.

\begin{figure}[t]
\centering
\hspace*{-0.8cm}
\includegraphics[width=.53\textwidth]{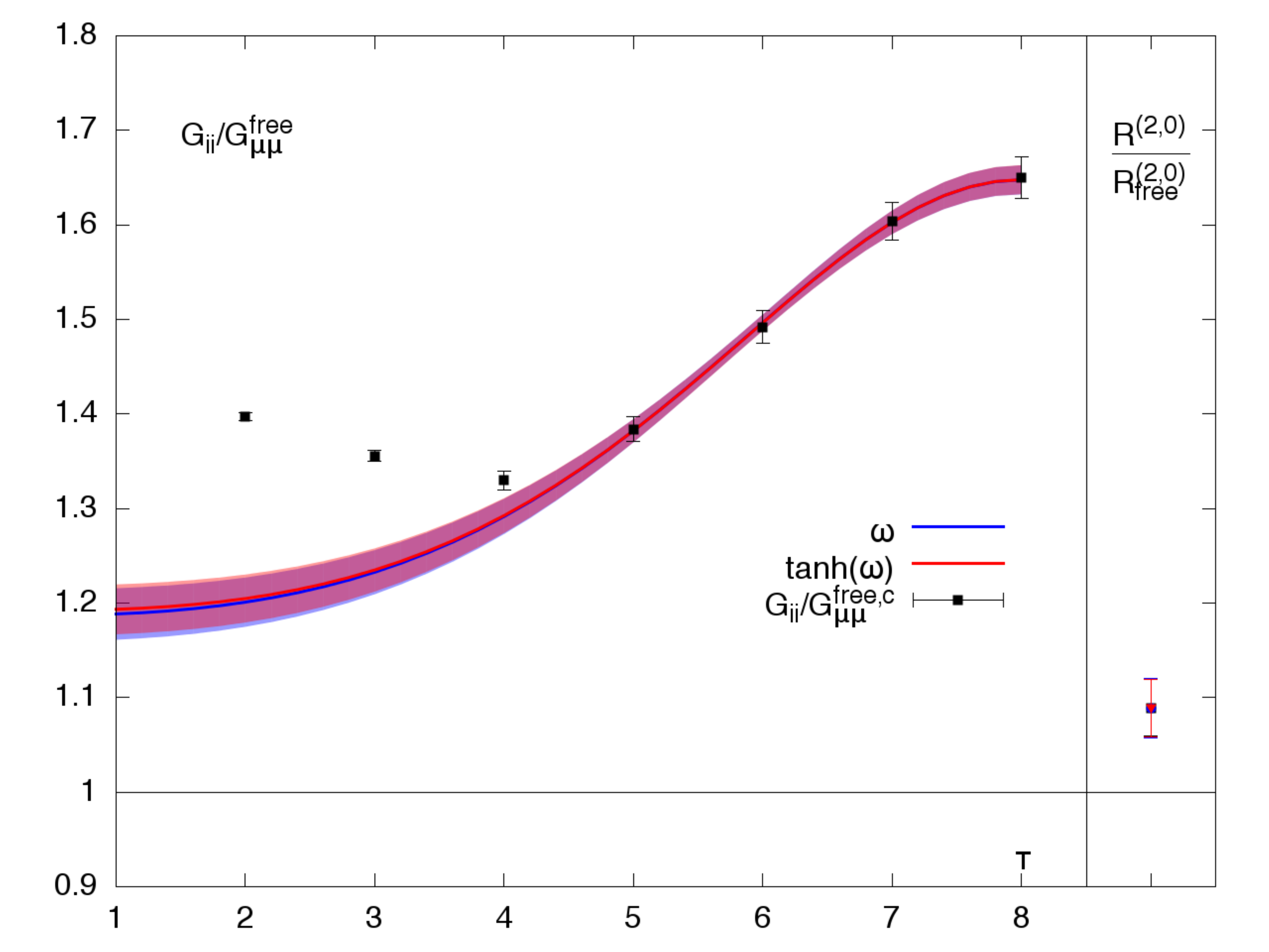}
\hspace*{-0.8cm}
\includegraphics[width=.53\textwidth]{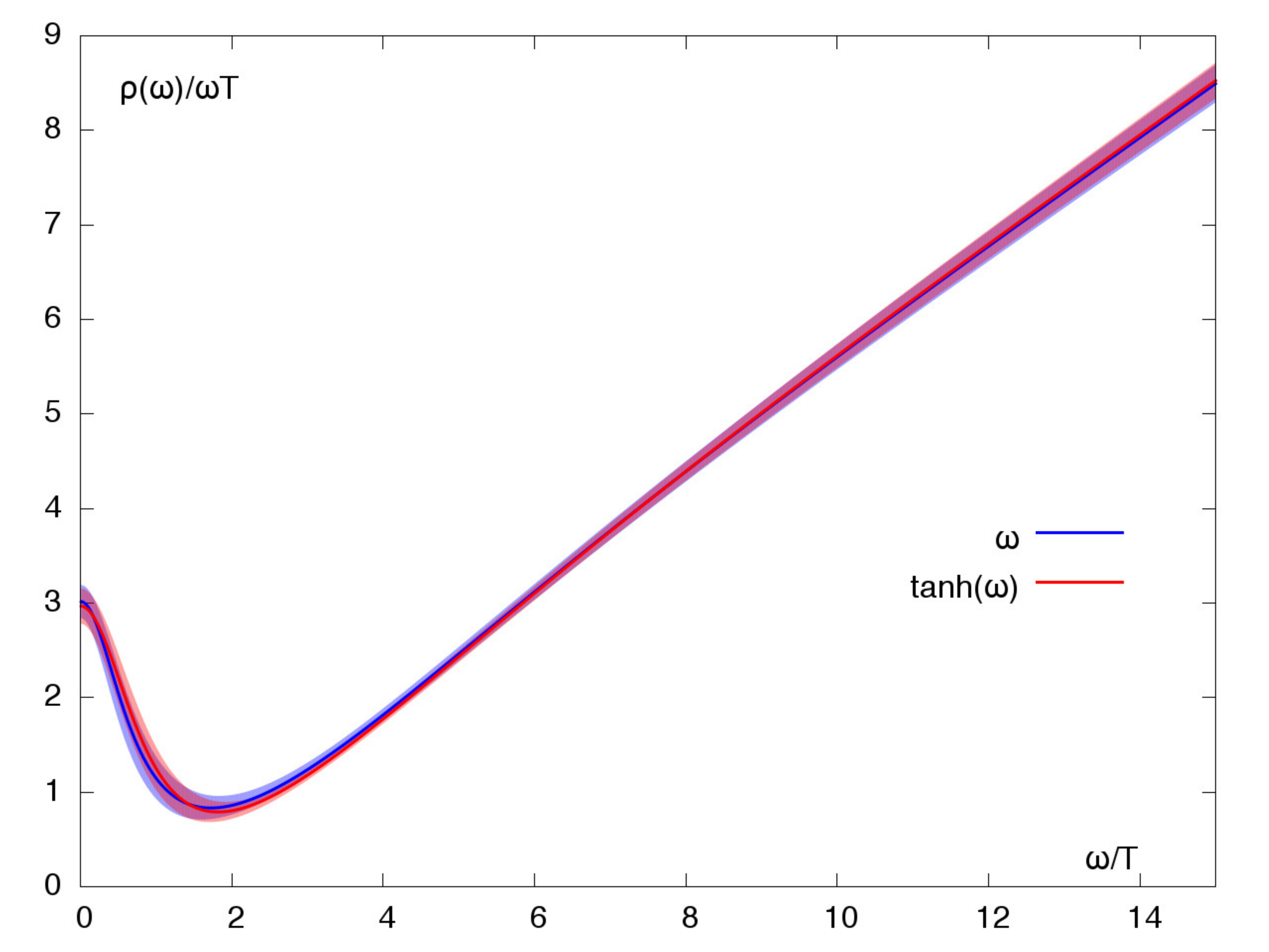}
\caption{\it\small{Left panel: fit results for $G_{ii}(\tau)/G_{\mu\mu}^{free}(\tau)$ 
using both Ans\"atze for the transport peak. 
 The resulting ratio of thermal moments
 $R^{(2,0)}/R^{(2,0)_{free}}$ are displayed on the right side of the plot.
 Right panel: The corresponding spectral functions normalized by $\omega T$.}} 
\label{fig:Dileptonfit}
\end{figure}

The resulting correlators  and spectral functions are shown in
Fig.~\ref{fig:Dileptonfit} with their fit parameters listed in
Tab.~\ref{tab:fits}. The ratio $G_{ii}(\tau)/G_{\mu\mu}^{\rm
  free}(\tau)$ in the left panel of Fig.~\ref{fig:Dileptonfit} is well
described by both versions $\rho_{T,1}$ and $\rho_{T,2}$ of the
transport contribution for $\tau/a\geq5$, while also the ratio of
thermal moments (given on the far right of the plot) is reproduced.
For $\tau/a<5$ our Ansatz fails to reproduce these points, which we
suspect is partly due to cutoff effects. In the future it would be
interesting to repeat the calculation at several smaller lattice
spacings while keeping the temperature fixed, as in
\cite{Ding:2010ga}.

On the right hand side of Fig.~\ref{fig:Dileptonfit} we show the
resulting spectral functions divided by $\omega T$.  Clearly both
Ans\"atze give very similar results that lie within errors of each
other. The thermal correlator is even less sensitive to the 
asymptotic behavior of the transport contribution in the Ansatz
than in the difference of correlators studied in Sec.~\ref{sec:GGrec}.

%%%%%%%%%%%%%%%%%%%%%%%%%%%%%%%%%%%%%%%%%%%%%%%%%%%%%%%%%%%%%%%%%%%%%%%%
%%%%%%%%%%%%%%%%%%%%%%%%%%%%%%%%%%%%%%%%%%%%%%%%%%%%%%%%%%%%%%%%%%%%%%%%
%%%%%%%%%%%%%%%%%%%%%%%%%%%%%%%%%%%%%%%%%%%%%%%%%%%%%%%%%%%%%%%%%%%%%%%%

\subsection{Discussion\la{sec:disc}}

\begin{table}
\centering
\begin{tabular}{llllll}
\hline\noalign{\smallskip}
$\Delta\rho(\omega,c,g,\kappa)$ & $c/T$ & $g/T$ & $c_B/T^3$  & $\mathcal{A}(\Lambda=1.5T)/T^2$ & $\langle v^2\rangle_{\rm eff}$\\
\noalign{\smallskip}\hline\noalign{\smallskip}
$\rho_{T,1}(\omega,c,g)$ & 0.61(10)          & 1.22(26)     & 1.42(188) &   0.702(201)  & 0.806(231) \\ 
$\rho_{T,2}(\omega,c,g)$ & 0.59(16)          & 5.7(10)      & 2.92(228) &   0.764(244)  & 0.877(280)\\ 
\hline\hline\noalign{\smallskip}
$\rho(\omega,c,g,\kappa)$ & $c/T$ & $g/T$ & $\kappa$  & $\mathcal{A}(\Lambda=1.5T)/T^2$ & $\langle v^2\rangle_{\rm eff}$ \\
\noalign{\smallskip}\hline\noalign{\smallskip}
$\rho_{T,1}(\omega,c,g)$ & 0.75(4)         &  0.71(9) &  1.186(27) &  0.818(50) & 0.939(57)\\     
$\rho_{T,2}(\omega,c,g)$ & 0.74(5)         & 0.98(13)     & 1.192(26) &  0.858(56) & 0.985(64)\\  
\noalign{\smallskip}\hline
\end{tabular}
\caption{\it\small{Parameters obtained from both studies. Top: fitting
$\Delta\rho(\omega,c,g,\kappa)$ 
to $\Delta G(\tau,T)$ with $g_B=0.5T$ and $\kappa=1.10$.
Bottom: fitting $G_{ii}(\tau)$ to
$\rho(\omega,c,g,\kappa)$. In both cases fits were done for two
different Breit-Wigner peaks in the low frequency region,
$\rho_{T,1}(\omega)\sim\omega$ and
$\rho_{T,2}(\omega)\sim T\tanh(\omega/T)$. Additionally the resulting area
of the transport region $\mathcal{A}(\Lambda)$ and the mean squared
velocity $\langle v^2\rangle$ is given. For details see the text.}}
\label{tab:fits}      
\end{table}

\begin{figure}[t]
\centering
\includegraphics[width=.63\textwidth]{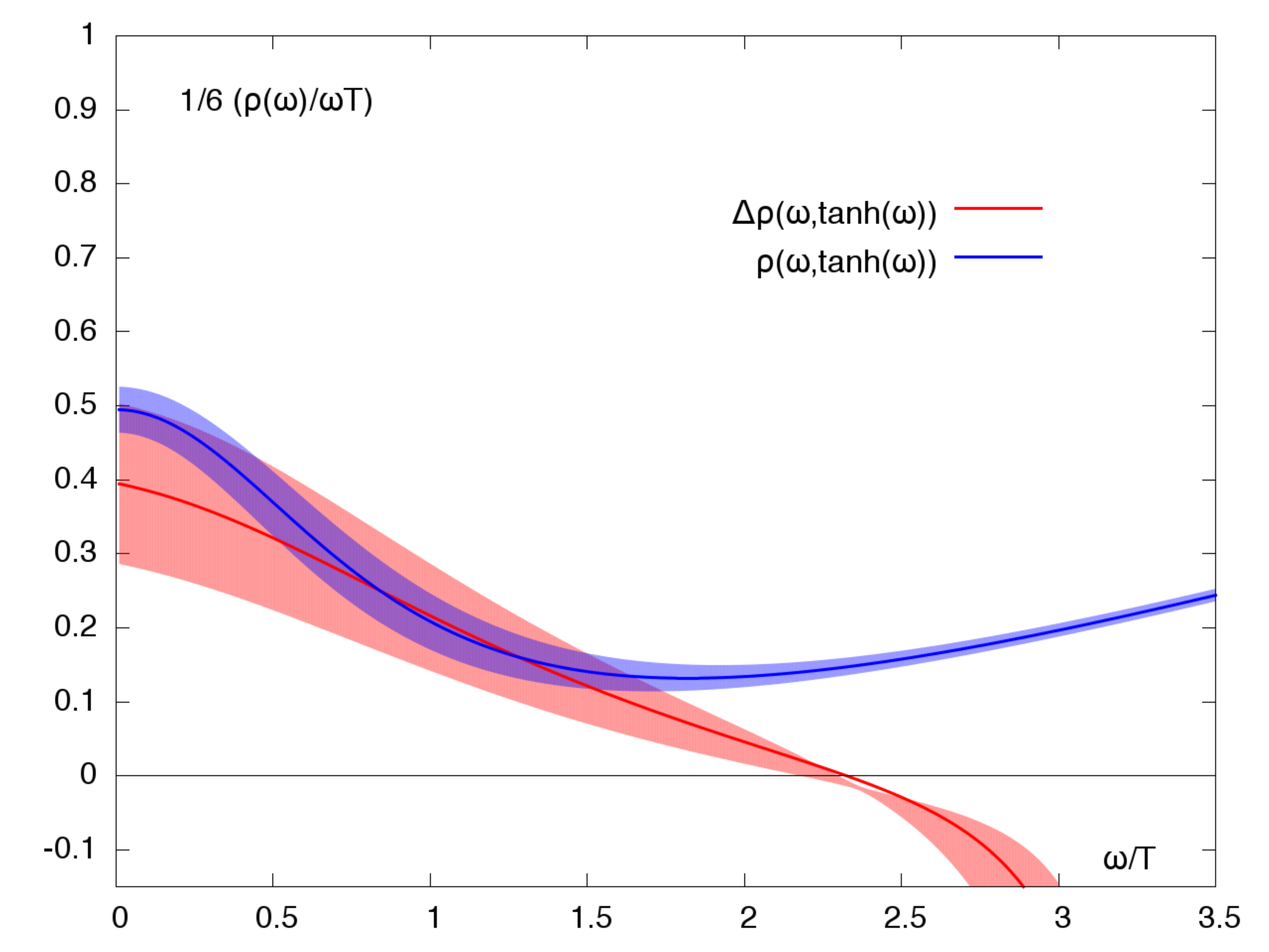}
\caption{\it\small{Comparison of the spectral functions obtained from analyzing
 (a) $\Delta G(\tau,T)$ and (b) $G_{ii}(\tau,T)$ 
using in both cases $\rho_{T,2}(\omega)\sim \tanh(\omega/T)$ in the low frequency region. 
All curves have been multiplied by a factor 1/6 and divided by $\omega T$, 
entailing that the intercept at $\omega=0$ yields an estimate of $\sigma/C_{em}T$.}}
\label{fig:Spectralfunctions}
\end{figure}

We now compare the results of the fits to $\Delta G(\tau,T)$ and
$G_{ii}(\tau,T)$. Since the vacuum spectral function vanishes below
$2m_\pi\approx 540{\rm MeV}$ (in infinite volume), $\rho_{ii}(\omega,T)$
and $\Delta\rho(\omega,T)$ should be equal for $\omega<2m_\pi\approx
2.1T$. We thus plot the spectral functions obtained from the two fits in
this frequency region, see Fig.~\ref{fig:Spectralfunctions}.  Here we
restrict ourselves to showing these results based on
$\rho_{T,2}(\omega,T)$, as their theoretical foundation is more sound
than those with $\rho_{T,1}(\omega,T)$.  All curves are multiplied by
a factor $1/6$, which means that the intercept at $\omega=0$ yields an
estimate of $\sigma/C_{em}T$ with $\sigma$ the electrical conductivity
of the quark gluon plasma.

The results obtained by fitting $G_{ii}(\tau,T)$ agree very well with
the central values obtained by fitting $\Delta G(\tau,T)$, as
summarized in Tab.~\ref{tab:fits}. The fit to $\Delta G(\tau,T)$ using
the transport Ansatz $\rho_{T,2}(\omega)$ yields a slightly lower
intercept.  If we assume the spectral function to be as smooth around
the origin as Fig.~\ref{fig:Spectralfunctions} suggests, we obtain the
following estimate for the electrical conductivity of the quark gluon
plasma at $T\simeq 250$MeV,
\be \la{eq:sigma_res} 
\frac{\sigma}{C_{em}T}= 0.40(12),
\ee 
where $C_{em} = \sum_{f=u,d} Q_f^2=5/9$ for $N_f=2$ and $C_{em}
=\sum_{f=u,d,s} Q_f^2=2/3$ for $N_f=2+1$, see Sec.~\ref{sec:pheno}.
It should be remembered that the Euclidean correlator can be perfectly
well described by an infinitely narrow transport peak (corresponding
to an infinite electrical conductivity), see section \ref{sec:simp}.
Although obtained under a strong assumption, it is interesting to
compare (\ref{eq:sigma_res}) to other lattice results obtained under
similar assumptions.  
The following comparison is made with quenched results, since 
to our knowledge there are no previous dynamical QCD studies.

A quenched calculation using staggered fermions based on fitting the
Fourier transform of the correlator obtained $\sigma/T=7C_{em}$
\cite{Gupta:2003zh} at $1.5\leq T/T_c\leq3.0$, where the pure SU(3)
gauge theory critical temperature is around 290$\,$MeV.  A further
study using staggered fermions and an analysis based on the maximum
entropy method obtained $\sigma/T=0.4(1)C_{em}$
\cite{Aarts:2007wj}. Finally, a recent quenched study using
Wilson-Clover fermions in the continuum limit obtained
$0.33C_{em}\leq\sigma/T\leq 1C_{em}$ at $T\simeq1.45T_c$
\cite{Ding:2010ga}. Our results using dynamical Wilson-Clover fermions
at $N_\tau=16$ are thus completely compatible with the recent quenched
results.

Even though the transport contribution extracted from our fits is not
narrow and an interpretation in terms of kinetic theory
(Eq.~\ref{eq:IntLbda}) is no longer rigorously motivated, we also
compute the effective mean squared velocity of the quarks $\langle
v^2\rangle$. Choosing the scale parameter $\Lambda/T=1.5$ and
numerically integrating the fitted spectral functions we obtain
$\mathcal{A}(\Lambda=1.5T)$. Assuming Eq.\ (\ref{eq:IntLbda}) and using
the quark number susceptibility as given in Tab.~\ref{tab:pars}, it is
straightforward to estimate an effective mean squared velocity
$\langle v^2\rangle_{\rm eff}$.  The results for
$\mathcal{A}(\Lambda=1.5T)$ and $\langle v^2\rangle_{\rm eff}$ are
listed in Tab.~\ref{tab:fits}. For all fits we obtain reasonable
values for $\langle v^2\rangle_{\rm eff}$ in the range
$0.80\lesssim\langle v^2\rangle_{\rm eff}\lesssim0.99$. Thus, although
we are not able to demonstrate or exclude the validity of the kinetic
theory description, the values of the effective quark velocity
extracted in this way are in line with its expectations.  By contrast,
the AdS/CFT spectral function (\ref{eq:N4vc}) (which clearly cannot be
described by kinetic theory) yields an effective quark velocity of
$\<v^2\>_{\rm eff} = 0.47$ for $\Lambda=1.5T$.

\begin{figure}[t]
\centering
\includegraphics[width=.63\textwidth]{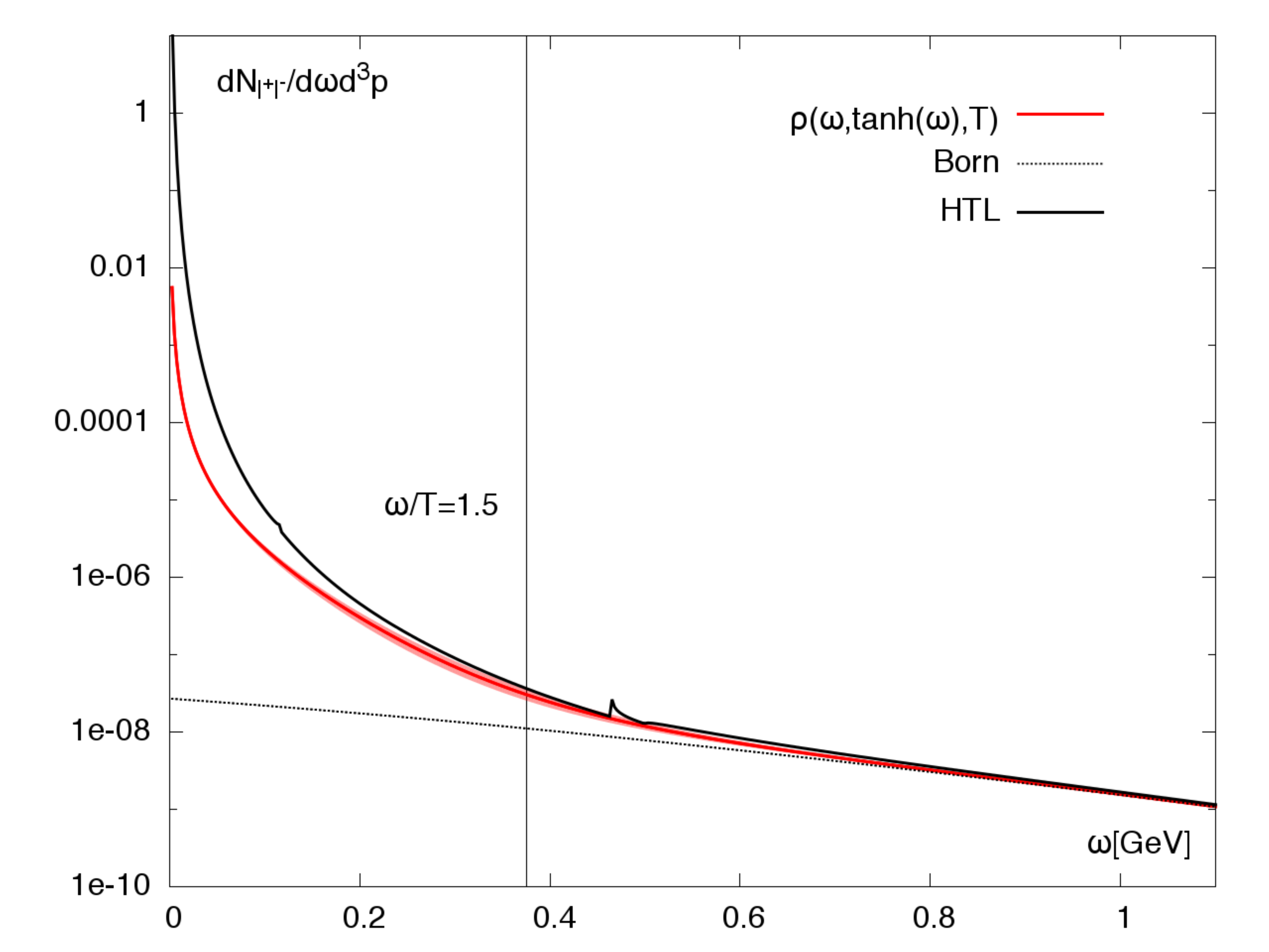}
\caption{\it\small{The production rate of lepton pairs from
two-flavor lattice QCD for frequencies $\omega$ given 
in units [GeV] as calculated from spectral functions of 
the vector current using Eq.~\ref{eq:dilrate}. The curve
comes from fitting $\rho_{ii}(\omega,T)$ to  $G_{ii}(\tau,T)$
using the transport peak Ansatz 2.
The black line shows the result from HTL perturbation theory 
with $m_T/T=1$, while the dashed line denotes the (free) Born rate. }} 
\label{fig:Dileptons}
\end{figure}

Using the result for $\rho_{ii}(\omega,T)$ from the fit to
$G_{ii}(\tau,T)$, it is straightforward to compute the production rate
of thermal lepton pairs in the quark gluon plasma with two light
dynamical quark flavors from Eq.\ (\ref{eq:dilrate}). The resulting rates
are shown in Fig.~\ref{fig:Dileptons}. We give the low frequency
behavior of $dN_{l^+l^-}/d\omega d^3p$ obtained from
$\rho_{ii}(\omega,T)$ using the Ansatz 2 in the transport region.
Comparing our results with the free (Born) rate (dashed line) and the
hard thermal loop result (black line) with a thermal mass of $m_T/T=1$
\cite{Braaten:1989mz}, we observe that for frequencies
$\omega/T\lesssim1.5$ our result is below the result from HTL, above
this value however it follows it very closely.

The results shown in Fig.~\ref{fig:Dileptons} correspond to a system
at thermal equilibrium with $T\simeq250$MeV.  To make contact with
results obtained in heavy ion collisions one has to take into account
the real-time evolution of the volume of the system, using a
hydrodynamic model along the lines of \cite{Rapp:2010sj} (for a recent
study of out-of-equilibrium photon and dilepton production using the
AdS/CFT correspondence see \cite{Baier:2012ax,Baier:2012tc}).

\section{Conclusion}

In this paper we have obtained the isovector vector correlator in the
high-temperature phase of two-flavor QCD at $T\simeq250$MeV as well as in
the vacuum at the same set of bare parameters. This allowed us to
analyze both the difference of the thermal and the vacuum correlator
and the thermal correlator directly. In the former case the analysis
is further constrained by an exact sum rule. Given the uncertainties
inherent in trying to extract information on the spectral function
from Euclidean correlators, the two methods give a consistent picture
of the thermal spectral function in the low to moderate frequency
range $\omega\lesssim 1.5T$.

The vacuum spectral function is known to receive a very large
contribution from the $\rho$ meson from experimental $e^+e^-$ and
$\tau$ decay data.  The main qualitative lesson we have learnt is that
the reduction or complete absence of such a peak and the appearance of
a substantial spectral weight in the low-frequency region provide a
very good description of the lattice data and are compatible with the
sum rule. Moreover the area under the latter spectral weight matches
the expectation of kinetic theory. This picture is corroborated by a
simple phenomenological study, presented in section
\ref{sec:phenorho}, based on the experimental $R(s)$ ratio and the sum
rule.  We also note that the analytic result (\ref{eq:N4vc}) in the
strongly coupled limit of ${\cal N}=4$ super-Yang-Mills theory
exhibits a qualitatively similar (but quantitatively different in
amplitude) change of sign in the difference of thermal and vacuum
spectral functions, even though the theory is conformal and therefore
exhibits no analogue of the $\rho$ meson.  A similar pattern was
also observed in the bulk channel of the pure SU(3) gauge
theory~\cite{Meyer:2010ii}.

The analysis presented in this paper is based on data at finite
lattice spacing.  Obviously the next step would be to repeat the
analysis in the continuum limit, as has been done in the quenched
theory~\cite{AnthonyThesis}. In this respect the results obtained here
should be regarded as preliminary. We note however that our results
are quantitatively quite close to those obtained in~\cite{AnthonyThesis}.

It would be very interesting to repeat the analysis carried out here
in a temperature scan through the smooth phase transition.  This would
allow one to track the fate of the $\rho$ meson from the
low-temperature to the high-temperature phase and perhaps to shed
light on the excess of dileptons observed by the PHENIX collaboration
in relativistic gold-gold collisions~\cite{Adare:2009qk}.  The methods
employed here to constrain thermal spectral functions may be useful in
the context of cosmology as well, see for
instance~\cite{Asaka:2006rw}.

\section*{Acknowledgments}
We are very grateful to Marina Marinkovic who generated the
zero-temperature ensemble used here that was made available to us
through CLS.  We also warmly thank Georg von Hippel who provided the vector
correlator on this ensemble.  H.M.\ thanks Aleksi Vuorinen for discussions.
We acknowledge the use of computing time for the generation of the gauge 
configurations on the JUGENE computer of the Gauss Centre
for Supercomputing located at Forschungszentrum J\"ulich, Germany,
allocated partly through the European PRACE initiative and 
partly through the John von Neumann Institute for Computing (NIC).
In particular, the finite-temperature ensemble 
was generated within NIC project HMZ21. The correlation functions were computed 
on the dedicated QCD platform ``Wilson'' at the Institute for Nuclear Physics,
University of Mainz. This work was supported by the
\emph{Center for Computational Sciences in Mainz} and by the DFG grant
ME 3622/2-1 \emph{Static and dynamic properties of QCD at finite
  temperature}.

\bibliographystyle{JHEP}
\bibliography{/home/meyerh/CTPHOPPER/ctphopper-home/BIBLIO/viscobib.bib}
\end{document}